%% file: Billaud_EMM-FM_2011_6.tex
\documentclass{article}

\usepackage[english]{babel}
\usepackage{graphicx, color, epsfig}
\usepackage{enumerate}
\usepackage{latexsym}
\usepackage{fancyhdr}
\usepackage{multirow}
\usepackage{cite}
\usepackage{amsmath,amssymb,amsthm}
\usepackage{ragged2e}

\newcommand{\ds}{\displaystyle}
\newcommand{\vs}{\vspace}

\newcommand{\A}{\mathbf{A}}
\renewcommand{\a}{\mathrm{a}}
\newcommand{\C}{\mathrm{C}}

\newcommand{\dd}{\mathrm{d}}
\newcommand{\ddd}{\mathbf{d}}
\newcommand{\E}{\mathbf{E}}
\newcommand{\e}{\mathrm{e}}
\newcommand{\ee}{\mathrm{e}}

\newcommand{\g}{\mathrm{g}}
\newcommand{\G}{\mathrm{G}}

\renewcommand{\j}{\mathrm{j}}

\newcommand{\h}{\mathrm{h}}

\newcommand{\mm}{\mathbf{m}}
\newcommand{\nn}{\mathbf{n}}
\newcommand{\p}{\mathbf{p}}
\newcommand{\PP}{\mathrm{P}}
\newcommand{\rr}{\mathbf{r}}
\newcommand{\rrr}{\mathrm{r}}

\newcommand{\Y}{\mathrm{Y}}

\newcommand{\sigmag}{\textrm{\mathversion{bold}$\sigma$}}

\newcommand{\ldc}{[\![}
\newcommand{\rdc}{]\!]}

\newcommand{\lambdabarre}{\!
\raisebox{.05cm}{$\begin{array}{c}
-
\\[-0.3cm]
\lambda
  \end{array}$}\!}

\newcommand{\lnd}{~\!\!&}
\newcommand{\rnd}{&\!\!~}

\newcommand{\Si}{\mathrm{Si}}

\title{\textbf{\LARGE Quantum properties of spherical semiconductor quantum dots}}

\author{B. \textsc{Billaud}\footnote{To whom correspondence should be addressed: \texttt{bbillaud@u-cergy.fr}}~~and T.-T. \textsc{Truong}
\ \\ \ \\
\textit{Laboratoire de Physique Th\'eorique et Mod\'elisation (LPTM),}
\\
\textit{CNRS UMR 8089, Universit\'e de Cergy-Pontoise,}
\\
\textit{2, av. Adolphe Chauvin, F-95302 Cergy-Pontoise Cedex, France.}}

\begin{document}
\maketitle

\abstract{Quantum effects at the nanometric level have been observed in many confined structures, and particularly in semiconductor quantum dots (QDs). In this work, we propose a theoretical improvement of the so-called effective mass approximation with the introduction of an effective pseudo-potential. This advantageously allows analytic calculations to a large extent, and leads to a better agreement with experimental data. We have obtained, as a function of the QD radius, in precise domains of validity, the QD ground state energy, its Stark and Lamb shifts. An observable Lamb shift is notably predicted for judiciously chosen semiconductor and radius. Despite the intrinsic non-degeneracy of the QD energy spectrum, we propose a {\it Gedankenexperiment} based on the use of the Casimir effect to test its observability. Finally, the effect of an electromagnetic cavity on semiconductor QDs is also considered, and its Purcell factor evaluated. This last result raises the possibility of having a QD-LASER emitting in the range of visible light.}
\ \\ \ \\
\par \begin{tabular}{rl}
{\it Keywords} & spherical Quantum Dot, semiconductor, exciton, Stark effect, Lamb effect,
\\
 & Casimir effect, Purcell Effect.
\\
{\it PACS} & 12.20.Ds, 71.35.-y, 71.70.Ej, 73.22.Dj.
  \end{tabular}
\section{Introduction}
Semiconductor quantum dots (QDs), as well as quantum wires or quantum wells, show properties of standard atomic physics, as a result of the restriction of the motion of one to a hundred conduction band electrons or valence band holes to a confined region of space of nanometric size. But, in contrast to atoms, phonons, surface effects and bulk disorder play a crucial role in determining their electronic properties, so that two QDs are never really identical. A QD may be thought as a giant artificial atom with an adjustable quantized energy spectrum, controlled only by its size. It enjoys prospects to serve in quantum optics as sources for a semiconductor LASER \cite{Kirkstaedter_1994} or of single photon \cite{Yoshie_2004}, in quantum information as {\it qubits} \cite{Trauzettel_2007}, in micro-electronics as single-electron transistors \cite{Ishibashi_2003}, or in biology and medicine as {\it fluorophores} \cite{So_2006}.

During the early 1980s, the so-called quantum size effects (QSE), characterized by a blue-shift of their optical spectra, has been observed in a large range of strongly confined systems \cite{Vojak_1980, Ekimov_1980, Golubkov_1981, Kash_1986, Temklin_1987}. It comes from a widening of the semiconductor optical band gap, due to the increase of the charge carriers confinement energy \cite{Brus_1984}. A review of empirical and theoretical results on quantum confinement effects in low-dimensional semiconductor structures is given in \cite{Yoffe_2002}. Modern approaches to this problem are discussed in \cite{Delerue, Bester_2009}. But, despite numerous theoretical and empirical models, to the best of our knowledge, there exists no simple and comprehensive one, which offers a significant analytic treatment. 

To apprehend the origin of QSE in spherical semiconductor QDs, we propose to adopt the effective mass approximation (EMA), which assumes parabolic valence and conduction bands \cite{Brus_1984, Efros_1982, Kayanuma_1988}. An electron and a hole behave as free particles with their usual effective mass, but confined in a spherical infinite potential well. Their Coulomb interaction is taken into account through a variational principle. This model, presented in section {\bf \ref{sec_2}}, allows the introduction of an effective pseudo-potential, which partially removes the characteristic overestimation of the electron-hole pair confinement energy for small QDs \cite{Thoai_1990}. As an achievement, an analytic expression for the phenomenological function $\eta(\lambda)$, introduced in \cite{Kayanuma_1988}, is obtained in good agreement with numerical data \cite{Billaud_2010a}.

Among many fundamental topics, the atom-like behavior of QDs is nowadays intensively investigated because of its potential technological applications. Of particular interest is the interaction with an external electromagnetic field. In semiconductor microcrystals, the presence of a constant electric field gives rise to quantum-confinement Stark effects (QCSE) \cite{Nomura_1990, Bawendi_1997, Patane_2000}. It manifests itself by a characteristic red-shift of the exciton photoluminescence \cite{Wen_1995, Yakimov_2003, Harutyunyan_2004, Ham_2005, Wei_2007}, and leads to a corresponding enhancement of its lifetime \cite{Polland_1985}. If an electric field is applied perpendicularly to the plane of multi-layers quantum wells, exciton energy shift peaks were measured and successfully compared to theoretical results \cite{Wood_1984}, obtained by a perturbative method introduced in \cite{Bastard_1983}, when the electron-hole Coulomb interaction is negligible. But, in spherical QDs, this turns out to be more important, and cannot be discarded \cite{Nomura_1990b}. In section {\bf \ref{sec_3}}, we propose to use the previous EMA model for spherical QDs. It allows the derivation of analytic criterions for choosing the QD radius and the applied electric field amplitude, as a result of the interplay between electron-hole Coulomb interaction and an additional polarization energy \cite{Billaud_2009}.

The Lamb shift in atoms, due to the interaction of valence electrons with  a quantized electromagnetic field, has provided a convincing experimental check of the validity of quantum electrodynamics, and has been, ever since, a continual subject of research. Effects of the band gap \cite{Sajeev_1990}, of a electromagnetic mode \cite{Brune_1994}, and its coupling to the QD surface \cite{Rotkin_2000} have been notably investigated in semiconductor QDs. However, the Lamb effect, which is experimentally well established \cite{Lamb_1947} and theoretically understood \cite{Bethe_1947, Welton_1948} in atoms by the end of the 1940s, seems to be unknown in QDs. The purpose of section {\bf \ref{sec_4}} is to fill this gap. The theoretical framework, set up in section {\bf \ref{sec_2}}, is used to evaluate the Lamb effect in a large range of spherical semiconductor QDs. In particular for small QDs, it can be shown that the electron-hole pair ground state undergoes an observable negative Lamb shift, at least for judiciously chosen semiconductors. Because of the intrisic non-degeneracy of QD energy levels, the problem of its experimental observability is put to question. A {\it Gedankenexperiment}, making use of the Casimir effect \cite{Casimir_1948}, is proposed to test its existence \cite{Billaud_2010b}.

To close this paper, the Purcell effect is investigated in section {\bf \ref{sec_5}}. This phenomenon is one of the striking phenomenon illustrating of cavity quantum electrodynamics \cite{Purcell_1946}. It consists of a significant enhancement of the spontaneous emission rate of quantum systems interacting with a resonant electromagnetic cavity mode \cite{Gerard_1998, Kiraz_2003}, which has found many applications, see {\it e.g.} \cite{Nakwaski_2003, Santori_2004}. Nowadays, it provides a test bed for quantum optics \cite{Imamoglu_2005} and quantum information \cite{Kiraz_2004}. A validity condition for obtaining Purcell effect in spherical semiconductor QDs is determined, in the presence of the adverse role of Rabi oscillations. Some predictive numerical results theoretically support the possibility of using the Purcell effect in such confined structures as radiative emitters in LASER devices.

A concluding section summarizes our main results and indicates some possible research directions.
\section{Quantum Size Effects} \label{sec_2}
In a standard EMA model, electrons and holes are assumed to be non-relativistic spinless particles, behaving as free particles with their effective masses $m^*_{\e,\h}$, in a confining infinite spherical potential well, written as, in spherical coordinates $(r,\theta,\varphi)$
  $$
V(\rr_{\e,\h})=\!
    \left\{
      \begin{array}{ccc}
0 & \textrm{if} & 0\leq r_{\e,\h}\leq R,
\\
\infty & \textrm{if} & r_{\e,\h}>R.
      \end{array}
    \right.\!\!
  $$
The choice of an infinite potential well at the QD surface induces an overestimation of the electron-hole pair ground state energy for small QDs, as compared to real finite potential. This can be usually corrected by restoring a finite potential step of experimentally acceptable height \cite{Thoai_1990}. However, the standard height of the realistic step potential implementing the confinement of the charge carriers inside the QD may be reasonably described by an infinite potential well. Electrons and holes are then isolated from the insulating surrounding of the QD. In this setting, as far as Stark, Lamb, Casimir or Purcell effects are concerned, the electromagnetic field amplitude should not exceed some threshold, so that electrons and holes would not acquire sufficient energy to overstep the real confining potential barrier by tunnel conductivity. We shall refer to this working assumption as the weak field limit.
\subsection{Interactive electron-hole pair EMA model} \label{subsec_2_1}
The Hamiltonian of an interactive electron-hole pair confined in a semiconductor spherical QD reads
  $$
H=H_\e+H_\h+V_\C(\rr_{\e\h}),
  $$
where, in units of $\hbar=1$, $H_{\e,\h}=-\frac{\nabla^2_{\!\e,\h}}{2m^*_{\e,\h}}+V(\rr_{\e,\h})$ denote the respective confinement Hamiltonians of the electron and of the hole, and $V_\C(\rr_{\e\h})=-\frac{e^2}{\kappa r_{\e\h}}$ the electron-hole Coulomb interaction, with $\kappa=4\pi\varepsilon$, $\varepsilon$ being the semiconductor dielectric constant, and $r_{\e\h}$ the electron-hole relative distance. Without loss of generality, the semiconductor energy band gap $E_\g$ may be set equal to be zero for convenience. In absence of Coulomb potential, electron and hole are decoupled particles with wave functions
  $$
\psi_{lnm}(\rr_{\e,\h})=\sqrt{\frac2{R^3}}\frac{\chi_{[0,R[}(r_{\e,\h})}{\j_{l+1}(k_{ln})}\j_l\left(\frac{k_{ln}}Rr_{\e,\h}\right)\Y^m_l(\theta_{\e,\h},\varphi_{\e,\h}),
  $$
where $\chi_\mathbb A(r)=\left\{
  \begin{array}{cl}
0 & \textrm{if}~~r\in\mathbb A
\\
1 & \textrm{otherwise}
  \end{array}
\right.$ is the radial characteristic function of the set $\mathbb A\subseteq\mathbb R_+$. $l\!\in\!\mathbb{N}$, $n\!\in\!\mathbb N\smallsetminus\{0\}$ and $m\!\in\!\ldc-l,l\rdc$ are quantum numbers, labeling the spherical harmonic $\Y^m_l(\theta,\varphi)$ and the spherical Bessel function of the first kind $\j_l(x)$. Finally, the wave numbers $k_{ln}$ are defined as the $n^{\textrm{\scriptsize th}}$ non-zero root of the function $\j_l(x)$, resulting from the continuity condition at $r=R$ \cite{Efros_1982}. The respective energy eigenvalues for electron and hole, expressed in terms of $k_{ln}$ as
  $$
E^{\e,\h}_{ln}=\frac{k_{ln}^2}{2m^*_{\e,\h}R^2},
  $$
show that the density of states of the semiconductor bulk has an atomic-like discrete spectrum, with increasing energy separation as the radius decreases. The analytical diagonalization of the Hamiltonian $H$ seems to be out of reach because the Coulomb potential explicitly breaks the spherical symmetry. To handle the interplay of the quantum confinement energy, scaling as $\propto R^{-2}$, and the Coulomb interaction, scaling as $\propto R^{-1}$, two regimes are to be distinguished, according to the ratio of the QD radius $R$ to the Bohr radius of the bulk exciton $a^*=\frac{\kappa}{e^2\mu}$, $\mu$ being the exciton reduced mass. In the strong confinement regime, corresponding to sizes $R\lesssim2a^*$, the electron-hole relative motion is sufficiently affected by the infinite potential well, so that {\it exciton} states should be considered as uncorrelated electronic and hole states. In the weak confinement regime, valid for sizes $R\gtrsim4a^*$, the exciton conserves its character of a {\it quasi}-particle of total mass $M=m^*_\e+m^*_\h$. Its center-of-mass motion is confined, and should be quantized \cite{Kayanuma_1988}.
\subsection{Strong confinement regime} \label{subsec_2_2}
In this regime, the Coulomb potential is treated as a perturbation with respect to the infinite confining potential well in a variational procedure. The ground state energy of the electron-hole pair can be evaluated with the trial wave function $\phi(\rr_\e,\rr_\h)=\psi_{010}(\rr_\e)\psi_{010}(\rr_\h)\phi_\mathrm{rel}(\rr_{\e\h})$, with $\phi_\mathrm{rel}(\rr_{\e\h})=\e^{-\frac\sigma2r_{\e,\h}}$, where $\sigma$ is the variational parameter. The product $\psi_{010}(\rr_\e)\psi_{010}(\rr_\h)$ insures that the confined electron and hole should both occupy their respective ground state in absence of Coulomb potential. The variational part $\phi_\mathrm{rel}(\rr_{\e\h})$ is chosen so that the electron-hole exhibits the behavior of an exciton bound state, analogous to the ground state of an hydrogen-like atom with mass $\mu$. Then, it is expected that $\sigma\propto a^{*-1}$. Integral representations for relevant diagonal matrix elements in the state $\phi(\rr_\e,\rr_\h)$ are analytically expressed in the Fourier transform formalism of relative electron-hole coordinates. To obtain the mean value of the Hamiltonian $H$, these expressions are Taylor-expanded with respect to the parameter $\sigma R$ near zero, up to the second order. This yields a value $\sigma_0=\frac{4B'}{a^*}$, for which the electron-hole energy is minimized\footnote{All constants appearing in the text and formulas are listed in Appendix {\bf\ref{appendix_A}}.}
  $$
E^\mathrm{strong}_{\e\h}=E_{\e\h}-A\frac{e^2}{\kappa R}-4B'^2E^*,
  $$
where $E_{\e\h}=E_\e+E_\h$, with the compact notation $E^{\e,\h}_{01}=E_{\e,\h}$, is the electron-hole pair ground state confinement energy, and $E^*=\frac1{2\mu a^{*2}}$ the binding exciton Rydberg energy. This formula has been already obtained with another trial function of the same form as $\phi(\rr_\e,\rr_\h)$, but with an interactive part equal to $\widetilde\phi_\mathrm{rel}(\rr_{\e\h})=1-\frac\sigma2r_{\e\h}$, instead of $\phi_\mathrm{rel}(\rr_{\e\h})$ \cite{Kayanuma_1988}, which obviously comes from the two first terms of the Taylor expansion of $\phi_\mathrm{rel}(\rr_{\e\h})$, in the limit of $\frac\sigma2r_{\e\h}\lesssim\frac R{a^*}\ll1$. Thus, this indicates that, taking first the limit $\frac R{a^*}\ll1$, and then evaluating matrix elements to perform the variational procedure, is equivalent to the reverse method applied here. This is not clear at first sight.
\subsection{Weak confinement regime}
In this regime, electron-hole pair states consist of exciton bound states. The Coulomb interaction contribution to the exciton ground state should no longer be considered as a perturbation to the confinement energy, but is still treatable as a perturbation to the infinite confining potential well. Therefore, the global form of the variational function $\phi(\rr_\e,\rr_\h)$ should be retained. However, the QD size allows a partial restoration of the long range Coulomb potential between the charged carriers, so that it is of the same order of magnitude than the kinetic energy in the electron-hole relative coordinates. Then, the leading contribution to the ground state energy of the exciton should be $-E^*$, the ground state energy of a hydrogen-like atom of mass $\mu$. The total translational motion of the exciton, thought as a {\it quasi}-particle of mass $M$, should be restored and contribute to the exciton total energy by an amount $\frac{\pi^2}{2MR^2}$, the ground state energy of a free particle trapped in a space region of typical size $R$. In a first approximation, the excitonic ground state energy is the sum of these two contributions. To improve phenomenologically its accuracy in regard to numerical simulations, a monotonic increasing function $\eta(\lambda)$ of the effective masses ratio $\lambda=\frac{m_\textrm{\tiny h}^*}{m_\textrm{\tiny e}^*}$ has been introduced in \cite{Kayanuma_1988}, as $E^\mathrm{weak}_{\e\h}=-E^*+\frac{\pi^2}{2M(R-\eta(\lambda)a^*)^2}$. Then, the exciton is preferentially thought as a rigid sphere of radius $\eta(\lambda)a^*$. Its center-of-mass, whose motion is quantized, could not reach the infinite potential well surface unless the electron-hole relative motion undergoes a strong deformation \cite{Kayanuma_1988}. 

To account for all these contributions, this suggests to multiply the trial function $\phi(\rr_\e,\rr_\h)$ by the ground state plane wave $\phi_\G(\rr_\G)=\ee^{i\frac\pi R\textrm{\scriptsize$\sigmag$}_\G\cdot\rr_\G}$, where $\rr_\G$ is the center-of-mass coordinates and $\sigmag_\G$ is a plane wave ground state quantum number vector satisfying the condition $|\sigmag_\G|^2=1$. The new trial function should then be $\psi(\rr_\e,\rr_\h)=\psi_{010}(\rr_\e)\psi_{010}(\rr_\h)\phi_\mathrm{rel}(\rr_{\e\h})\phi_\G(\rr_\G)$, leaving unchanged the exciton density of probability as well as the Coulomb potential matrix element. The confinement Hamiltonian $H_\e+H_\h$ mean value gets the expected further contribution $\frac{\pi^2}{2MR^2}$. A Taylor expansion on the Hamiltonian $H$ mean value is performed in the region of QD radii $\sigma R\gtrsim2\pi$, and the value of the variational parameter $\sigma_0\approx\ds2a^{*-1}$ is then computed. Its second and third order terms in $\frac{a^*}R\lesssim1$ are neglected, because they do not contribute to the total electron-hole pair ground state energy, up to the third order in $\frac{a^*}R$,
  $$
E^\mathrm{weak}_{\e\h}=-E^*+\frac{\pi^2}{6\mu R^2}+\frac{\pi^2}{2M(R-\eta(\lambda)a^*)^2}, 
  $$
from where an analytical expression for the function $\eta(\lambda)$ can be extracted
  \begin{equation}
\eta(\lambda)=\delta\frac{(1+\lambda)^2}\lambda. \label{eta}
  \end{equation}
This obviously satisfies the electron-hole exchange symmetry $\lambda\rightarrow\lambda^{-1}$. If $l^*=\frac32a^*$ denotes the electron-hole relative distance mean value in the non-confined exciton ground state, the smallest possible radius $\frac12l^*=\frac34a^*$ for the excitonic sphere picture should be obtained when $\lambda=1$, so that $\eta(1)\approx\frac34$. This matches quite well with both numerical and theoretical results, as shown in table \ref{table_1}. We can compare values from numerical simulations taken by the function $\eta(\lambda)$ for $\lambda=1,3,5$ from \cite{Kayanuma_1988} to those theoretically predicted by Eq. (\ref{eta}), and observe that there exists a reasonable agreement between both results. In the same spirit, the largest possible radius $l^*$ should be reached in the infinite hole mass limit $\lambda\rightarrow\infty$, because the hole is motionless and located at the electron-hole system center-of-mass. According to table \ref{table_1}, $\eta(5)\approx\frac32$, it is reasonable to conclude that Eq. (\ref{eta}) is valid so long as $\lambda\lesssim5$, whereas in the infinite hole mass limit is reached for $\lambda\gtrsim5$, and $\eta(\lambda)\approx\frac32$.

  \begin{table}
\caption{\label{table_1}Comparison of $\eta(\lambda)$ values from numerical results of \cite{Kayanuma_1988} and theoretical ones given by Eq. (\ref{eta}).}
    \begin{center}
      \begin{tabular}{cccc}
\hline\hline
$\lambda$ & 1 & 3 & 5
\\
\hline
$\eta_\mathrm{num}(\lambda)$ & 0.73 & 1.1 & 1.4
\\
$\eta_\mathrm{theo}(\lambda)$ & 0.83 & 1.1 & 1.5
\\
relative error & $\approx$14\% & $<$1\% & $\approx$7\%
\\
\hline\hline
      \end{tabular}
    \end{center}
  \end{table}
\subsection{A pseudo-potential-like method}
When compared with \cite{Kayanuma_1988}, the exciton ground state energy $E^\mathrm{weak}_{\e\h}$ shows a further contribution $\frac{\pi^2}{6\mu R^2}$, which should be interpreted as a kinetic energy term in the relative coordinates because of the reduced mass $\mu$. As the virial theorem in this set of coordinates should be satisfied, this energy is already contained in the Rydberg energy term, and should be removed from $E^\mathrm{weak}_{\e\h}$. To this end, we propose to introduce an additional potential $W(\rr_{\e\h})$ to the electron-hole pair Hamiltonian $H$, which should make contributions to the second order of the exciton total energy in the weak confinement but not to the third one, the one responsible for the expression of the function $\eta(\lambda)$. In this picture, higher order contributions are interpreted as higher order corrections to $\frac{\pi^2}{2MR^2}$. This reinforces the idea that for very large radius, only the {\it quasi}-particle point of view should be responsible for the exciton kinetic energy. Such potential is uniquely determined to be of the form
  $$
W(\rr_{\e\h})=-\frac{32\pi^2}9E^*\frac{r_{\e\h}^2}{R^2}\ee^{-2\frac{r_{\e,\h}}{a^*}},
  $$
while the amplitude of $W(\rr_{\e\h})$ is to be fixed to get the correct kinetic energy $-\frac{\pi^2}{6\mu R^2}$, up to the third order in $\frac{a^*}R$, in the weak confinement regime. It is attractive at distances $\approx a^*$ to promote excitonic state with typical size around its Bohr radius, repulsive at short distances to penalize excitonic state with small size, and exponentially small for large distances not to perturb the long range Coulomb potential. Finally, it contributes to the exciton ground state energy with second order terms in both weak and strong confinement regimes, but does not change its zeroth and first order terms.

\begin{figure}
\caption{Ground state energy of an interactive electron-hole pair, confined by an infinite potential well with (--$\!~$--) or without (---) the presence of the pseudo-potential $W(\rr_{\e\h})$ and by a finite potential step of height $V_0\approx1$eV (--$~\!\cdot\!~$--) \cite{Thoai_1990} and compared to experimental results for spherical CdS microcrystals \cite{Weller_1986}.} \label{figure_1}
    \begin{center}
\input{figure_1.pstex_t}
    \end{center}
  \end{figure}
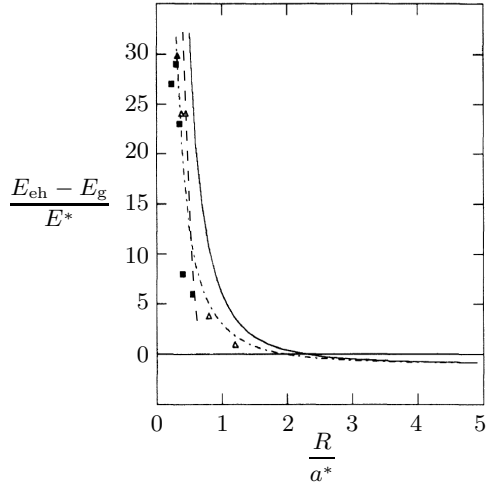

The addition of the pseudo-potential $W(\rr_{\e\h})$ to the exciton Hamiltonian $H$ implies a significant decrease of the expected value of the exciton energy in the strong confinement regime by an amount $-\frac{64\pi^2}9CE^*$, up to the second order in $\frac R{a^*}$. However, this is only valid when $R\lesssim \frac12a^*$ because of the pseudo-potential exponential dependence. Figure \ref{figure_1} shows that the excitonic energy computed in presence of the pseudo-potential gets a better fit to experimental results in this validity domain, than those calculated without this tool. Nevertheless, the divergence for very small QD size still persists as a consequence of the infinite potential well assumption. To extend the validity domain, energy expansions may be carried out to a few orders. But, calculations become so involved that the relevance of such an approach cannot be ascertained.

\section{Quantum-confinement Stark effects} \label{sec_3}
The model discussed in the previous section lends itself to an extensive amount of analytical calculations on spherical semiconductor nanostructures interacting with a fixed external electric field. Even if this model has some intrinsic limitations and does not fully describe the QD behavior in the absence of electric field for small QD radii, it can be still used, since it gives rather satisfactory theoretical predictions on Stark effect in the weak field limit.

Contrary to the case of a large range of microstructures, in which QCSE significantly depend on the electric field direction \cite{Bastard_1983, Spector_2005, Ham_2005, Wei_2007, Spector_2007, Pacheco_1997}, the applied electric field $\E_\a$, in spherical QDs, is set along the $z$-direction of a cartesian coordinates system with its origin at the QD center. As the inside semiconductor QD dielectric constant $\varepsilon$ is larger than the outside insulating matrix dielectric constant $\varepsilon'$, the electric field $\E_\dd$ inside the QD, different from $\E_\a$, is $\E_\dd=\frac3{2+\varepsilon_\rrr}\E_\a$ \cite{Wen_1995}. Then, the electron and the hole interaction Hamiltonians with the electric field $\E_\dd$ are
  \begin{eqnarray}
W_{\e,\h}(\rr_{\e,\h})=\pm e\E_\dd\cdot\rr_{\e,\h}=\pm eE_\dd r_{\e,\h}\cos\theta_{\e,\h},
  \end{eqnarray}
where $E_\dd$ is the electric field amplitude inside the microcrystal. As $\langle\phi|W_\e(\rr_\e)|\phi\rangle=-\langle\phi|W_\h(\rr_\h)|\phi\rangle$, in presence of an electric field, a new dependence on the electron and hole space coordinates for the trial wave function is required. The difference between the  dielectric constants also implies the existence of a polarization energy term
  $$
P(\rr_\e,\rr_\h)=\frac{e^2}{2R}\sum_{l\geq1}\frac{\alpha_l(\varepsilon_\rrr)}{R^{2l}}\left(r_\e^{2l}+r_\h^{2l}-2r^l_\e r^l_\h\PP_l(\cos\theta_{\e\h})\right),
  $$
where $\PP_l(x)$ denotes a Legendre polynomial, $\varepsilon_\rrr$ the relative dielectric constant, and $\alpha_l(\varepsilon_\rrr)=\frac{(l-1)(\varepsilon_\rrr-1)}{\kappa(l\varepsilon_\rrr+l+1)}$ \cite{Brus_1984}. The polarization energy $P(\rr_\e,\rr_\h)$ will be neglected first, but taken into account later on, to explore in details its relative role {\it vs.} the Coulomb potential.
\subsection{Variational wavefunction} \label{subsec_3_1}
To apprehend QCSE, we follow the reasoning of \cite{Bastard_1983}, and study the interaction between the charge carriers with the ambient electric field but neglecting their Coulomb interaction. In the weak field limit, the absolute value of their interaction energy with the electric field $\E_\dd$ is $E_\mathrm{ele}=eE_\dd R$. It should be treated as a perturbation compared to their typical confinement energy $E_{\e,\h}$. The Stark shift will be computed by perturbation and variational procedures on individual Hamiltonians $H'_{\e,\h}=H_{\e,\h}+W_{\e,\h}(\rr_{\e,\h})$. The trial function to be used is $\Phi_{\e,\h}(\rr_{\e,\h})=\psi_{010}(\rr_{\e,\h})\varphi_{\e,\h}(\rr_{\e,\h})$, where $\varphi_{\e,\h}(\rr_{\e,\h})=\e^{\mp\frac{\sigma_{\e,\h}}2r_{\e,\h}\cos\theta_{\e,\h}}$, because it contains a deformation of the spherical shape along the electric field direction. The minimizing variational parameters $\sigma_{\e,\h}$ are found to be
$\sigma^0_{\e,\h}=\frac{4C}3m^*_{\e,\h}eE_\dd R^2$.

Both methods lead to the Stark shift $-\Gamma m_{\e,\h}^*e^2E_\dd^2R^4$, where the proportionality coefficients are $\Gamma_\mathrm{pert}=\frac{32\pi^2}3\sum_{n\geq1}\frac{k_{1n}^2}{(k_{1n}^2-\pi^2)^5}$ and $\Gamma_\mathrm{var}=\frac{2C^2}9$, with a relative error of about $2\%$. Since, the variational function $\phi(\rr_\e,\rr_\h)$ describes the electron-hole pair Coulomb interaction, both occupying their respective ground state, and the electric field interaction part $\varphi_\e(\rr_\e)\varphi_\h(\rr_\h)$ is liable for the individual electron and hole behaviors in the electric field $\E_\dd$, this suggests to choose a variationnal function in presence of the electric field of the form $\Phi(\rr_\e,\rr_\h)=\phi(\rr_\e,\rr_\h)\varphi_\e(\rr_\e)\varphi_\h(\rr_\h)$.
\subsection{Stark effect in absence of polarization energy} \label{subsec_3_2}
To describe Stark effects in spherical semiconductor microcrystals without taking into account polarization effects, we apply a variational procedure using the trial function $\Phi(\rr_\e,\rr_\h)$ to the Hamiltonian
  $$
H_\mathrm{Stark}=H+W_\e(\rr_\e)+W_\h(\rr_\h).
  $$
Fourier transform techniques lead to integral representation of diagonal matrix elements, valid if and only if the variational parameters $\sigma$ and $\sigma_{\e,\h}$ satisfy the inequality $0\leq\ee\cdot\sigma_{\e,\h}<\sigma$, where $\ee=\exp(1)$. This relation analytically expresses the range of acceptable electric field amplitudes. Following previous results of sections {\bf\ref{subsec_2_2}} and {\bf\ref{subsec_3_1}}, we respectively expect that $\sigma\propto a^{*-1}$ and $\sigma_{\e,\h}\propto m_{\e,\h}^*eE_\dd R^2$, so that $E_\mathrm{ele}\propto\frac{\sigma_{\e,\h}}\sigma \frac R{a^*}E_{\e,\h}$. Then, the charge carriers energy due to their interaction with the electric field should be at the most of the same order of magnitude of a first term correction in $\frac R{a^*}$ to their confinement energy in the strong confinement regime. This corresponds to the order of magnitude of the typical absolute electron-hole Coulomb interaction. Based on the decoupled electron-hole point of view, the Stark shift for the coupled electron-hole pair should scale as $\propto (m_\e^*+m_\h^*)e^2E_\dd^2R^4\propto E_{\e\h}\frac{R^2}{a^{*2}}$. Therefore, to get at least the lowest order contribution to this Stark shift, a Taylor expansion of the Hamiltonian $H_\mathrm{Stark}$ mean value is performed up to the second order in the variational parameters. This first contribution is not sufficiently accurate to fit experimental data, because it does not account for the electron-hole coupling through the Coulomb interaction. This is the reason why the expansion up to the third order should be carried out to obtain the first correction in $\frac R{a^*}$.\footnote{For later purpose, let us give the expression of the Taylor expansion of the Coulomb potential mean value
  $$
\frac{\langle\Phi|V_\C(\rr_{\e\h})|\Phi\rangle}{\langle\Phi|\Phi\rangle}=-\frac{e^2}{\kappa R}\left(A+B'\sigma R+C'\sigma^2R^2+C'_1(\sigma_\e^2+\sigma_\h^2)R^2+C'_2\sigma_\e\sigma_\h R^2+O(\sigma^3R^3)\right).
  $$}
Then, the Stark shift, being identified as with the term scaling as $\propto E_\dd^2$, is determined, up to the first order in $\frac R{a^*}$, as\footnote{The values of the variational parameters are also obtained up to the first order in $\frac R{a^*}$
  \begin{eqnarray*}
\sigma'_0\lnd=\rnd4B'\left(1+8C'\frac R{a^*}\right)-\frac83CC''(m^*_\e+m^*_\h)\frac{e^2E_\dd^2R^4}{E^*}\frac R{a^*},
\vs{.2cm}
\\
\sigma_{\e,\h}^0\lnd=\rnd\frac{4C}3m_{\e,\h}^*eE_\dd R^2\left[1+4\left(2C'_1\frac{m^*_{\e,\h}}\mu+C'_2\frac{m^*_{\e,\h}}{\mu}-\frac{3B'C''}C\right)\frac R{a^*}\right].
  \end{eqnarray*}}
  $$
\Delta E^\mathrm{strong}_\mathrm{Stark}=-\Gamma_\mathrm{var}(m^*_\e+m^*_\h) e^2E_\dd^2R^4\left(1+8\Gamma^{\e\h}_\mathrm{var}\frac R{a^*}\right),
  $$
where $\Gamma_\mathrm{var}$ appears as a universal constant, while the constant $\Gamma^{\e\h}_\mathrm{var}$ depends on the semiconductor, {\it i.e.}
  $$
\Gamma^{\e\h}_\mathrm{var}=C'_1\left(\frac{m^*_\e}{m^*_\h}+\frac{m^*_\h}{m^*_\e}\right)+C'_2-\frac{3B'C''}C.
  $$
Furthermore, this model is capable of describing QCSE, when the QD size and the electric field amplitude satisfy effective constraints, consistent with strong confinement regime and weak field limit
  $$
\frac R{a^*}\lesssim\frac1{2(3B'+4C')}~~~~\textrm{and}~~~~\frac{E_\mathrm{ele}}{E_{\e\h}}\lesssim\frac1{\pi^2\ee C\left(1+\frac43\frac{C'}{B'}\right)}.
  $$
As expected, the first contribution to the found shift is simply the sum of the Stark shift contributions undergone by the ground states of both electron and hole taken individually. Because of the dependence of $\Gamma^{\e\h}_\mathrm{var}$ on the effective masses $m^*_{\e,\h}$, the second contribution to $\Delta E^\mathrm{strong}_\mathrm{Stark}$ indicates the existence of a dipolar interaction between the electron and the hole. Until now, the interaction between the electron or the hole with the external electric field takes place individually, whereas they interact only through the Coulomb potential. Actually, the Hamiltonian interaction part should also be written as $W_{\e\h}(\rr_{\e\h})=W_\e(\rr_\e)+W_\h(\rr_\h)=\E_\dd\cdot\ddd_{\e\h}$, where $\ddd_{\e\h}=e\rr_{\e\h}$ is the exciton electric dipole moment. In the strong confinement regime, the dipolar interaction point of view expresses the remnant of electron-hole pair states, thought as exciton bound states under the influence of the electric field.
\subsection{Stark effect in presence of polarization energy} \label{subsec_3_3}
The dipolar interaction point of view suggests the inclusion of the term $P(\rr_\e,\rr_\h)$ in the Hamiltonian $H_\mathrm{Stark}$ describing the exciton-electric field interaction \cite{Brus_1984}, which accounts for the polarization energy of the electron-hole pair, due to the difference between the dielectric constants of the semiconductor QD and its insulating surrounding. A variational procedure is applied to the new Hamiltonian $H'_\mathrm{Stark}=H_\mathrm{Stark}+P$. For this, we keep the variational trial function $\Phi(\rr_\e,\rr_\h)$, since the polarization energy should not basically modify the nature of the electron-hole coupling. The polarization energy mean value is expressed as an expansion in the variational parameters, which has the same form as the Coulomb potential mean value, where the constants $A$, $B'$, $C'$, $C'_1$ and $C'_2$ are replaced by functions of the relative dielectric constant $\varepsilon_\rrr$, as shown in table \ref{table_7}.\footnote{The polarization energy mean value in the state $\Phi(\rr_\e,\rr_\h)$ should be written as
  \begin{equation}
\frac{\langle\Phi|P(\rr_\e,\rr_\h)|\Phi\rangle}{\langle\Phi|\Phi\rangle}=\frac{-e^2}{\kappa R}\left\{A(\varepsilon_\rrr)+B'(\varepsilon_\rrr)\sigma R+C'(\varepsilon_\rrr)\sigma^2R^2+C'_1(\varepsilon_\rrr)(\sigma_\e^2+\sigma_\h^2)R^2+C'_2(\varepsilon_\rrr)\sigma_\e\sigma_\h R^2+O(\sigma^3R^3)\right\}. \label{Phi|P(rr_e,rr_h)|Phi}
  \end{equation}}
In this formalism, expressions for any Stark effect quantity, pertaining either to polarization energy or to the combined effect of Coulomb interaction and polarization energy are obtained from section {\bf \ref{subsec_3_2}}. All the appearing constants are replaced either by the corresponding functions of $\varepsilon_\rrr$ or by the sum of both contributions.
\subsection{Comparison with experimental data}
\begin{figure*}
\caption{Stark shift for confined interactive electron-hole pair only including the Coulomb interaction up to the zeroth (---) or to the first (--$\!~$--) order, with $\Gamma^{\e,\h}_\mathrm{var}\approx-0.1629$; including both the Coulomb interaction and the polarization energy up to the zeroth (---) or to the first order (--$\!~$--), with $\Gamma^{\e,\h}_\mathrm{var}\approx-0.2045$, and including only the polarization energy up to the first order (--$\!~\cdot\!~$--), with $\Gamma^{\e,\h}_\mathrm{var}\approx-0.0416$, in comparison with experimental results (+) \cite{Nomura_1990} in spherical CdS$_{0.12}$Se$_{0.88}$ microcrystals.} \label{figure_2}
    \begin{center}
\input{figure_2.pstex_t}
    \end{center}
  \end{figure*}
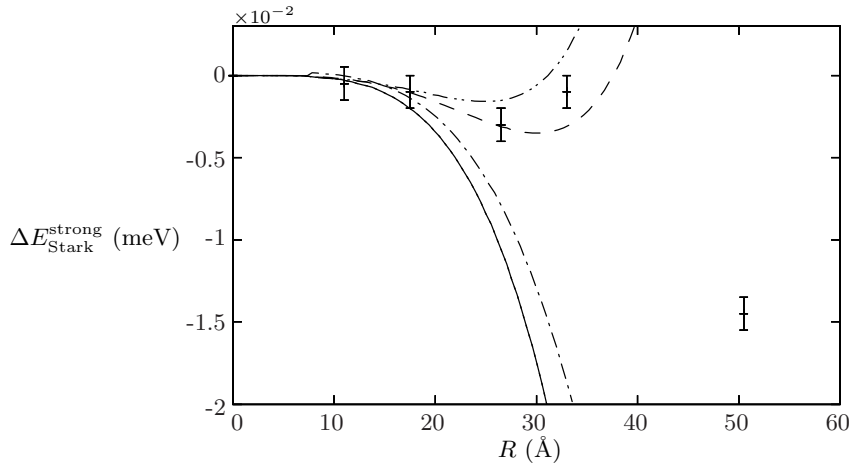

Figure \ref{figure_2} shows our theoretical predictions {\it vs.} experimental data for spherical CdS$_{0.12}$Se$_{0.88}$ microcrystals \cite{Nomura_1990}. The electric field amplitude inside the microcrystal is set at $E_\dd=12.5$kV.cm$^{-1}$. Two exciton peaks are experimentally resolved, which are attributed to the transitions from the highest valence sub-band and from the spin-orbit split-off state to the lowest conduction sub-band, with an energy splitting about 0.39eV, independently of the QD radius \cite{Nomura_1990}. The experimental values depicted by crosses in figure \ref{figure_2} consist of mean values of the Stark shift of these two types of excitons. They seem to indicate that the Coulomb interaction is sufficient to explain correctly the amplitude of the Stark effects experimentally observed, as we expect, in the range of validity of QD radii.

In the case of CdS$_{0.12}$Se$_{0.88}$ microcrystals, our predictions should lead to acceptable results in regard to experimental data as long as the cluster radius does not exceed 30\AA~if only the Coulomb interaction is taken to account, or 50\AA~if the polarization is also included. When only polarization is considered, the strong confinement regime is no longer valid, because $\sigma'_0\leq0$. This means that polarization energy is repulsive and the interactive part of the trial function should be $\phi_\mathrm{rel}(r_{\e\h})=\ee^{|\sigma'_0|r_{\e\h}}$. In the three previous study cases, the hypothesis of the weak electric field limit remains valid as soon as the typical electric dipole interaction energy $E_\mathrm{ele}$ does not represent about 12\% of the typical exciton confinement energy $E_{\e\h}$. The weak field limit seems to be judiciously chosen, because it appears to be independent of the strong confinement regime.

When only the Coulomb potential is taken into account, the highest acceptable electric field amplitude, consistent with the weak field limit, is $E^\mathrm{max}_\dd\approx450$kV.cm$^{-1}$ for $R=10$\AA, and $E^\mathrm{max}_\dd\approx16.7$kV.cm$^{-1}$ for $R=30$\AA. Thus, these values show that, for $E_\dd\approx12.5$kV.cm$^{-1}$ and for $R\lesssim30$\AA, the strong confinement regime and the weak field limit are satisfied.

Figure \ref{figure_2} shows that the absolute value of the Stark shift, computed up to the zeroth order, is significantly overestimated, except for very small QD radii. The results become much more accurate, if the first order is included. This seems to be efficient enough to describe QCSE in spherical semiconductor QDs for $R\lesssim30$\AA. As soon as $R\gtrsim30$\AA, our results diverge significantly from experimental data.

When polarization energy is included, the same orders of magnitude in $E^\mathrm{max}_\dd$ are involved. Whereas the strong confinement regime condition is satisfied, for QD radii 30\AA~$\lesssim R\lesssim50$\AA, the weak field limit is no longer valid. This may explain the significant divergence from experimental results in this region. Figure \ref{figure_2} also clearly shows that the behavior of the polarization energy, if considered alone, is not satisfactory. If the Coulomb potential is taken into account, the results accurately fit the experimental data, except if $R\approx30$\AA, for which the weak field limit breaks down.  Thus, a future research work may focus on considering the strong confinement regime in the limit of strong electric field --- or even more generally to any electric field amplitude. The case of the weak confinement regime of the electron-hole pair is much more difficult, even in the weak field limit. An entirely different approach may be needed.
\section{Lamb effect and its possible observability} \label{sec_4}
According to the Dirac theory, the levels $2s$ and $2p$ of the hydrogen atom should be degenerated. But, they are actually experimentally split by an energy of about 0.033cm$^{-1}$ \cite{Lamb_1947}. This phenomenon consists of the so-called Lamb shift. It has been generally attributed to the potential $V(\rr)$ undergone by a spinless massive particle of mass $m^*$ and charge $qe$ in interaction with a quantized dynamical electromagnetic field, which is well described by the Pauli-Fierz Hamiltonian in the Coulomb gauge
  $$
H_\mathrm{PF}=H_\mathrm{free}+H_\mathrm{em}+eH_\mathrm{int},
  $$
where $\A(t,\rr)$ is the electromagnetic potential vector \cite{Pauli_1938}. Here, $H_\mathrm{free}=\frac{\p^2}{2m^*}+V(\rr)$ is the {\it free} particle Hamiltonian, whose eigenvectors and eigenvalues are $|\nn\rangle$ and $E_\nn$, labeled by the quantum numbers $\nn$. $H_\mathrm{em}$ is the Hamiltonian of the electromagnetic field, on which the standard second quatization method is to be applied. Since experimental light sources possess sufficiently weak intensities, the weak field regime is valid, and the potential vector quadratic terms are discarded ahead of the linear term in $H_\mathrm{PF}$, so that the interaction Hamiltonian is expressed as $H_\mathrm{int}=-q\frac{\A\cdot\p}{m^*}$.
\subsection{General considerations}
There exists two methods to correctly apprehend the Lamb shift effect. The Bethe approach is a perturbation procedure applied to the Hamiltonian $H_\mathrm{PF}$ with respect to the perturbative Hamiltonian $H_\mathrm{int}$ \cite{Bethe_1947}. The Welton approach interprets the Lamb shift as a fluctuation effect \cite{Welton_1948}. Using standard arguments from quantum electrodynamics, they both lead to the same general expression for the Lamb shift of the particle state $|\nn\rangle$, given in terms of potential Laplacian matrix elements
  \begin{equation}
\Delta E_\nn=\frac\alpha{3\pi}\frac{q^2}{m^{*2}}\log\left(\frac{m^*}{\kappa^*}\right)\langle\nn|\nabla^2V(\rr)|\nn\rangle, \label{DelatE_lamb}
  \end{equation}
where $\alpha$ is the fine-structure constant, and $\kappa^*$ an IR cut-off, which is identified with the mean value of all level differences absolute values $\langle|E_\mm-E_\nn|\rangle$ \cite{Welton_1948}. The Bethe approach historically allowed the theoretical explaination of the Lamb shift for hydrogen \cite{Bethe_1947}. Even if this two different methods produce the same predictive result, the Welton approach brings a deeper comprehension for the Lamb shift as a physical phenomena, and satisfies to an invariant gauge property. Whereas, the Bethe perturbative argument attributes the Lamb shift to a weak radiation-matter couplage in Coulomb gauge.

This is the reason why we present in more details the Welton approach. It consists of a semi-classical point of view, in which the position of a quantum particle fluctuates around its mean position $\rr$ with small fluctuations $\Delta\rr$, due to its interaction with a classical surrounding electromagnetic field. The mean square oscillation amplitude position of a charged particle coupled to the zero-point fluctuations of the electromagnetic field can be easily evaluated as $\langle(\Delta\rr)^2\rangle=\frac{2\alpha}\pi\frac{q^2}{m^{*2}}\log\left(\frac{m^*}{\kappa^*}\right)$. In this picture, $\kappa^*$ is interpreted as the minimal wave pulsation, for which some particle position fluctuations should be observed, and could be determined {\it a posteriori} by considerations on the particle classical motion. Position fluctuations lead to a modification of the potential, whose the mean value over an isotropic distribution of the fluctuations $\Delta\rr$ should be computed. This gives rise to a mean effective potential $\langle V(\rr+\Delta\mathbf{r})\rangle=\left\{1+\frac{\langle(\Delta\rr)^2\rangle}6\nabla^2+\dots\right\}V(\rr)$.
The second order correction term $\Delta V(\rr)=\frac{\langle(\Delta\rr)^2\rangle}6\nabla^2V(\rr)$ is responsible for the Lamb shift, as its mean value $\Delta E_\nn$ in a state $|\nn\rangle$ gives the energy shift of Eq. (\ref{DelatE_lamb}).
\subsection{Lamb effect for a confined particle}
Let us consider the previous massive particle with $q=\pm 1$ confined by a spherical infinite potential well $V(\rr)$ described by eingenfunctions $\psi_{lnm}(r,\theta,\varphi)$ and energy eingenvalues $E_{ln}=\frac{k^2_{ln}}{2m^*R^2}$. To use Eq. (\ref{DelatE_lamb}), the infinite potential well Laplacian matrix element $\langle\psi_{lnm}|\nabla^2V(\rr)|\psi_{lnm}\rangle$ and the IR cut-off $\kappa^*$ in the spherical infinite potential well are to reckon. The laplacian of $V(\rr)$ cannot be evaluated even in distributions formalism. All calculations are then made with an intermediate finite potential step of height $V$, and the limit $V\rightarrow\infty$ is taken at last. The Lamb shift undergone by a state $|\psi_{lnm}\rangle$ of the particle confined by the infinite potential well $V(\rr)$ is assumed to be the finite part (only contribution independant from $V$) of the expansion in powers of $V$ of the Lamb shift undergone by the state of the particle confined by the finite potential step, with same quantum numbers $l$, $n$ and $m$. By construction, it seems that the IR cut-off $\kappa^*=\langle |E_{ij}-E_{ln}|\rangle$ is infinitely large if all possible quantum numbers are taken into account. This is not the case, because the confined particle interacts with the surrounding electromagnetic field, which possesses a finite energy. It is not appropriate to consider that the particle can access all its energy levels states. There exists a higher energy level that it can attain by its interaction with the electromagnetic field, which consists of the total electromagnetic energy $E_\mathrm{lim}$. This allows to define a associated maximal pulsation $\kappa_\mathrm{lim}$ by $E_{ \textrm{\scriptsize lim}}=\frac{\kappa_{\textrm{\tiny lim}}^2}{2m^*R^2}$, interpreted as a UV cut-off for the autorized wave numbers $k_{ln}\leq\kappa_\mathrm{lim}$. 

These considerations lead to expressions for the Lamb shift and the IR cut-off
  \begin{equation}
\frac{\Delta E_{ln}}{E_{ln}}=-\frac{16\alpha}{3\pi}\frac{\lambdabarre^{*2}}{R^2}\log\left(\frac R{R^*_\mathrm{min}}\right),~~~~\textrm{and}~~~~\kappa^*(R)=\frac{7\pi^2}{12m^*R^2},\label{DE_lamb_general}
  \end{equation}
where $\lambdabarre^*=m^{*-1}$ is the reduced particle Compton wavelength. Due to the potential well spherical symmetry, it is expected that the Lamb shift is independent of the azimuthal quantum number $m$. The radius $R^*_\mathrm{min}=\frac\pi2\sqrt{\frac73}\lambdabarre^*\approx2.399\lambdabarre^*$ insures the validity of the non-relativistic point of view. If $R\leq R^*_\mathrm{min}$, the particle should acquire a confinement energy at least of the same order of magnitude that its mass energy $m^*$. If the confined particle is an electron in vacuum, the Lamb shift is not measurable for reasonable potential well sizes. This will not be the case, if we consider a confined interactive electron-hole pair in spherical semiconductor QDs, at least in the strong confinement regime.

The description of the confined electron-pair made in section {\bf \ref{sec_2}} is integrated to the Welton approach to Lamb shift. Then, the Lamb shift of the exciton ground state presents four contributions of different kinds. The first two contributions are due to the confinement infinite potential well $V(\rr_{\e,\h})$ given by Eqs. (\ref{DE_lamb_general}), where the electron and hole effective masses are used. The second is due to the Coulomb potential $V_\C(\rr_{\e\h})$, and is of the same nature that the Lamb shift observed in real atoms. And, the third comes from the pseudo-potential $W(\rr_{\e\h})$.
\subsection{Lamb effect in spherical semiconductor QDs}
Thanks to Welton approach to the Lamb effect coupled to our description of QSE, we achieve to determine analytical expressions for the Lamb shift undergone by the ground state of a electron-hole pair confined in a spherical semiconductor QD.

In the strong confinement regime, we deduce, up to the second order in $\frac R{a^*}$, that the Lamb shift undergone by the electron-hole pair ground state should be written as
  $$
\Delta E^\mathrm{strong}_\mathrm{Lamb}=\Delta E^\mathrm{strong}_\e+\Delta E^\mathrm{strong}_\h,
  $$
where 
  $$
\frac{\Delta E^\mathrm{strong}_{\e,\h}}{E_{\e\h}}=-\frac{16\alpha}{3\pi\varepsilon}\frac{\lambdabarre^{*2}_{\e,\h}}{R^2}\log\left(\frac R{R^{\e,\h}_\mathrm{min}}\right)\left[1-\left(\frac{\mu F}{m^*_{\e,\h}}+\frac{2\mathfrak A}{\pi^2}\right)\frac R{a^*}+\left(\frac{\mu F'}{m^*_{\e,\h}}-\frac{2F''}{\pi^2}+\frac83\right)\frac{R^2}{a^{*2}}+O\left(\frac{R^3}{a^{*3}}\right)\right]\leq0,
  $$
and $\lambdabarre^*_{\e,\h}=m^{*-1}_{\e,\h}$ and $R^{\e,\h}_\mathrm{min}=\frac\pi2\sqrt{\frac73}\lambdabarre^*_{\e,\h}$ are respectively the electron and the hole reduce Compton wavelengths and minimal radii in the considered semiconductor. Since by definition $R\geq R^{\e,\h}_\mathrm{min}$, this Lamb shift is negative. This is an outstanding, property predicted for the first time, to the best of our knowledge, because in real atoms, Lamb effect always raise energy levels. Heuristically, the quantized electromagnetic field non-zero ground state energy, often called the zero-point energy, is responsible for spreading of the charge and mass of the carriers in a sphere of a typical radius $\sqrt{\langle(\Delta \rr)^2\rangle}$. This is the major phenomenon to the Lamb effect. In atoms, these fluctuations induce a screening of the Coulomb potential, resulting in a reduction of the binding energy of the electron to the nucleus \cite{Eides_2001}. The situation is different in a QD, the observed effect is not due to an electric charge spreading but to a mass spreading. The total energy, initially concentrated in the kinetic energy of the point-like particle, is now transferred to the energy of a spatial mass distribution, which splits into center of mass motion and relative motion. On the basis of total energy conservation, a reduction of the center mass motion energy is then expected, an effect which is opposite to the one observed in atoms.
  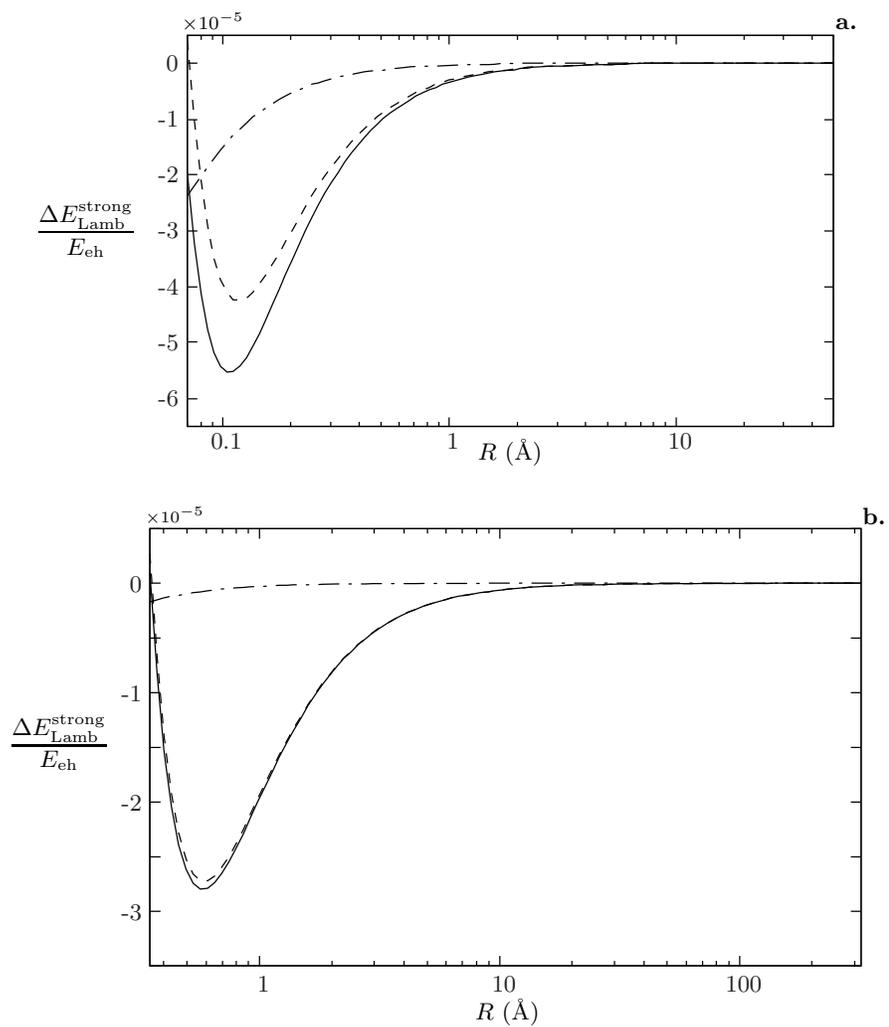
\begin{figure}
\caption{Lamb shift undergone by the electron (--$\!~$--), the hole (--$\!~\cdot~\!$--) and the exciton (---) in the strong confinement regime {\bf a.} in spherical CdS$_{0.12}$Se$_{0.88}$ or {\bf b.} in spherical InAs (heavy hole) microcrystals.} \label{figure_3}
    \begin{center}
\input{figure_3a.pstex_t}
\\ \ \\
\input{figure_3b.pstex_t}
    \end{center}
  \end{figure}

In the weak confinement regime, following a same reasoning, the Lamb shift of the exciton ground state is determined, up to the third in $\frac{a^*}R$, as
  $$
\Delta E^{\mathrm{weak}}_\mathrm{Lamb}=\Delta E^{\mathrm{weak}}_\e+\Delta E^{\mathrm{weak}}_\h,~~~~\textrm{where}~~~~\frac{\Delta E^{\mathrm{weak}}_{\e,\h}}{E^*}=\frac{8\alpha}{3\pi\varepsilon}\frac{\lambdabarre^{*2}_{\e,\h}}{a^{*2}}\log\left(\frac{m_{\e,\h}^*}{\kappa^*_{\e,\h}}\right)\left[1+O\left(\frac{a^{*3}}{R^3}\right)\right].
  $$
This energy shift is independant from the QD raduis, and reveals the excitonic {\it quasi}-particle properties of the electron-hole pair in the weak confinement regime. In the limit of infinite hole mass, only the electronic term $\Delta E^{\mathrm{weak}}_\e$ contributes to the Lamb shift of the exciton ground state, so that the Lamb shift undergone by the ground state of an hydrogen-like atom of reduce mass $\mu$ and Bohr raduis $a^*$ is retrieved. Following this analogy, the IR cut-offs $\kappa^*_{\e,\h}$ are both taken as $\kappa^*_{\e,\h}\approx19.8E^*$, because the IR cut-off associated to the hydrogen atom ground state yields $\approx19.8E_I$, where $E_I\approx13.6$eV is the hydrogen ionization energy \cite{Bethe}. Let us point out here the important fact that the contribution of the pseudo-potential $W(\rr_{\e,\h})$ plays a significant role on the Lamb effect of the exciton ground state in the weak confinement regime. Contributions due to the Coulomb interaction to the second order in $\frac{a^*}R$ are exactly discarded by contributions due to the pseudo-potential, the presence of the infinite potential well affecting the exciton Lamb shift in the weak confinement regime only by contributions of at least the fifth order. This supports the phenomenological introduction of the pseudo-potential $W(\rr_{\e,\h})$.

  \begin{table}
\caption{Lamb shift undergone by the exciton ground state in the strong confinement regime in spherical CdS$_{0.12}$Se$_{0.88}$ or InAs microcrystals {\bf a.} for $R=10$\AA~and {\bf b.} for $R=30$\AA, and {\bf c.} in the weak confinement regime.} \label{table_2}
    \begin{center}
      \begin{tabular}[t]{ccccc}
\hline
\hline
 & \multirow{2}{13ex}{Semiconductor} & \multirow{2}{13ex}{CdS$_{0.12}$Se$_{0.88}$} & \multicolumn{2}{c}{InAs}
\\
 &  &  & heavy hole & light hole
\\
\hline
{\bf a.} & $\Delta E^\mathrm{strong}_\mathrm{Lamb}$ ($\mu$eV) & -2.05 10$^{-2}$ & -9.49 & -36.9
\vs{.1cm}
\\
{\bf b.} & $\Delta E^\mathrm{strong}_\mathrm{Lamb}$ ($\mu$eV) & -2.11 10$^{-3}$ & -0.148 & -0.594
\vs{.1cm}
\\
{\bf c.} & $\Delta E^\mathrm{weak}_\mathrm{Lamb}$ ($\mu$eV) & 7.25 10$^{-3}$ & 2.69 10$^{-7}$ & 9.07 10$^{-5}$
\vs{.1cm}
\\
\hline
\hline
      \end{tabular}
    \end{center}
  \end{table}

Figure \ref{figure_3} show this Lamb shift for CdS$_{0.12}$Se$_{0.88}$ and InAs microcrystals, and suggest the possibility of observing it. More precisely, table \ref{table_2} confirms that the energy order of magnitude involved in the strong confinement regime in InAs microcrystals are equivalent to those in hydrogen atom. The observation of this Lamb shift in the weak confinement regime seems to be out of question for the moment, since exciton Rydberg energies in semiconductors are at most of the order magnitude of ten or so meV. We can wisely conclude that, at least in the strong confinement regime and in a judiciously chosen semiconductor, it seems possible to observe Lamb effect.
\subsection{Observability of the Lamb effect in spherical semiconductor QDs}
The experimental observability of the Lamb effect in hydrogen atom is possible because the $s$-spectral band is separated from $p$-spectral band, while they should stay degenerated in absence of Lamb effect. In quantum systems displaying no spectral band degeneracy, such as QDs, energy levels are dressed by the quantum zero-point fluctuations of the electromagnetic field, forbidding the detection of the corresponding bare levels, and then of the Lamb shift. In quantum field theory, the summation of the zero-point energy yields a divergent ground state energy, which is usually subtracted off in an additive renormalization scheme. However, a careful analysis on its boundary conditions shows the occurrence of a finite and observable force, known as Casimir force \cite{Casimir_1948, Mohideen_1998}. In vacuum, two parallel perfectly conducting squared plates of linear size $L$, placed at a separation distance $d\ll L$, are subjected to an attractive force, since the vacuum fluctuations are more important outside than inside the plates.

The Lamb effect is also attributed to the zero-point fluctuations energy of the electromagnetic field. So, by placing a QD in vacuum and inside a Casimir pair of conducting plates, one would be able to detect an energy difference between two Lamb shifted levels. This {\it Gedankenexperiment} should allow to overcome the need of degenerate energy levels, or of exactly computed energy levels. There exist some theoretical works dealing with Lamb effect of real atoms confining in a Casimir device \cite{Cheon_1988, Jhe_1991}. They predict an additional shift to the standard Lamb shift, which depends on the separation distance between the mirrors, which goes to zero in the limit $d\rightarrow\infty$. The coupling between the atom and its own radiation field is usually neglected. This assumption should be valid if the coupling of a two-level quantum atom with itself through absorption and emission of dipolar radiations reflected by the Casimir plates is dominated by the coupling of the two-level atom with the electromagnetic field vacuum fluctuations. For spherical semiconductor QDs in strong confinement regime, it means that $\kappa_d=\frac\pi d\leq\kappa^*_{\e,\h}(R)$. A direct generalization of these works in such context is possible, and leads to the addition of a new positive term $\Delta E^\mathrm{Casimir}_{\e,\h}(d)$ to the Lamb shift undergone by the confined electron-hole pair in a Casimir configuration in comparison with the one in vacuum
  $$
\Delta E^\mathrm{strong}_{\e,\h}\longrightarrow\Delta E^\mathrm{strong}_{\e,\h}+\Delta E^\mathrm{Casimir}_{\e,\h}(d),~~~~\textrm{where}~~~~\frac{\Delta E^\mathrm{Casimir}_{\e,\h}(d)}{\Delta E^\mathrm{strong}_{\e,\h}}=\frac6{49\pi^2}\!\left(\frac{R^2}{\lambdabarre^*_{\e,\h}d}\right)^2\log^{-1}\left(\frac R{R^{\e,\h}_\mathrm{min}}\right)\geq0.
  $$

This additional positive term calls for a physical explanation. When the separation distance $d$ decreases, the amplitude of the electromagnetic modes inside the Casimir plates increases, while their number is fixed. This leads to the reinforcement of the interaction of the quantum system with the quantized electromagnetic field, implying a strengthening of the Lamb effect. Moreover, this relative enhancement does not depend on the quantum state under consideration. Finally, there is a competition between the typical lengths of the QD: $\lambdabarre^*_{\e,\h}\ll R\ll d$, describing the different scales of the problem.

\begin{figure}
\caption{Modification of the Lamb shift in spherical InAs (light hole) microcrystals for $d=1\mu$m (---), $0.5\mu$m (--$\!~$--) or $0.25\mu$m (--$\!~\cdot~\!$--).} \label{figure_4}
    \begin{center}
\input{figure_4.pstex_t}
    \end{center}
  \end{figure}
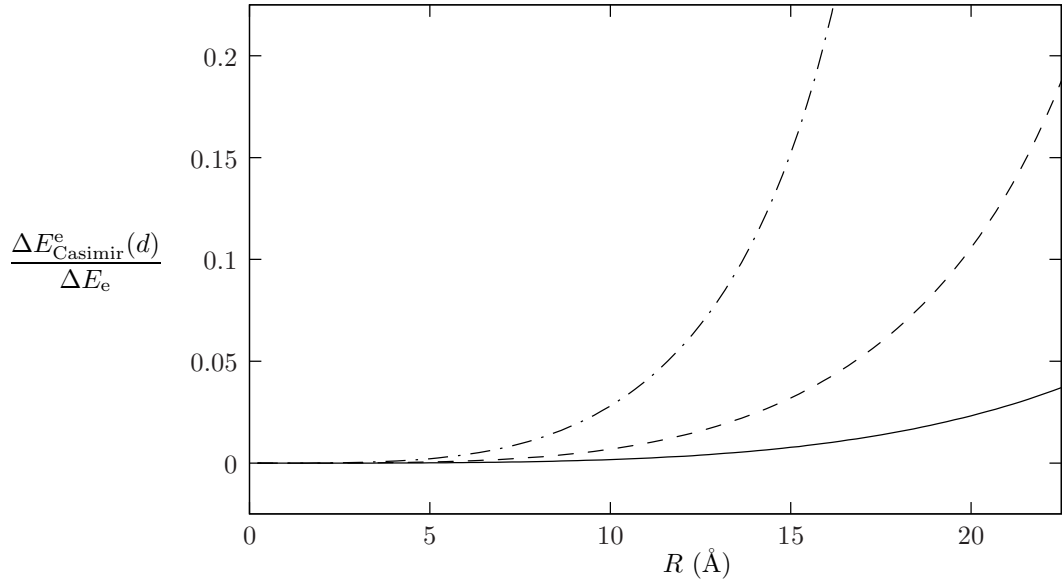

If the separation distance $d$ is chosen to be of 0.5$\mu$m, such that it allows the experimental observation of the Casimir effect, figure \ref{figure_4} shows that modification of the electron-hole Lamb shift between the Casimir plates is of about 5-10\% in spherical InAs QDs of radius of the order of magnitude of a few tens nm. It is possible to enhance the amplitude of this modification, of course, by reducing the separation distance $d$ until the order of a few tenth parts $\mu$m, or more simply by acting on the Casimir configuration geometry. For example, the use of a sphere of large radius instead of one of the Casimir plates increases the Casimir force, and then the modification of the Lamb shift, by a factor $\pi$ \cite{Mohideen_1998}. The combination of these two effects almost leads to the doubling of the Lamb shift in free space, which seems significant enough to be observable.
\section{Purcell effect} \label{sec_5}
It has been observed that coupling a magnetic moment to a resonating circuit of volume $V$ and quality factor $Q$ at radio frequencies of wavelength $\lambda$ significantly enhances its spontaneous emission by a factor $F=\frac3{4\pi^2}\frac{Q\lambda^3}V$ \cite{Purcell_1946}. This effect can be understood in a simple way. A two-level quantum atom, built from two eigenstates $|\nn\rangle$ and $|\mm\rangle$ of a quantum charged particle Hamiltonian $H_\mathrm{free}$, with respective energy eigenvalues $E_\nn<E_\mm$, is fit into a resonant electromagnetic cavity at a frequency $\omega$ close to the Bohr frequency $\omega_\nn^\mm=E_\mm-E_\nn$, with a quality factor $Q$. It interacts with a single dynamical confined electromagnetic cavity mode, also named {\it quasi}-mode, characterized by its effective volume $V_\mathrm{mode}$. In this picture, the {\it quasi}-mode is not only coupled to the two-level quantum atom but also to the {\it continuum} of other electromagnetic field modes. To compute spontaneous emission rates, the coupling between the two-level quantum system and the {\it quasi}-mode is treated as a perturbation when compared to the coupling between the confined mode and the {\it continuum} of external modes. As shown in \cite{Fano_1961}, in this weak coupling regime, the confined mode is characterized by a normalized Lorentzian energy density distribution $\rho(E)=\frac{2Q}{\pi\omega}\frac1{4Q(1-\frac E\omega)^2+1}$, of width $|\omega-\omega_\nn^\mm|=\frac\omega Q$. In the electric dipole approximation, the perturbation Hamiltonian is the standard dipolar interaction Hamiltonian
  $$
W(t,\rr)=-\ddd\cdot\E(\rr)\Theta(t),
  $$
where $\Theta(t)$ is the Heaviside step function, $\ddd$ the particle  dipole moment, and $\E(\rr)$ the quantized {\it quasi}-mode electric field, which reaches its maximal amplitude at the origin of a cartesian coordinates system.
\subsection{General considerations}
The general spontaneous emission transition rate $A_\nn^\mm$ associated to the radiative transition $|\mm\rangle \rightarrow |\nn\rangle$ with emission of a photon of pulsation $\omega\approx\omega_\nn^\mm$ is given by the Fermi golden rule between the two quantum states $|\mm,0\rangle$ and $|\nn,1\rangle$ of the Jaynes-Cummings Hamiltonian \cite{Jaynes_1963}, describing the two-level quantum atom-{\it quasi}-mode coupled system,
  $$
H_\mathrm{JC}=H_\mathrm{free}+H_\mathrm{em}+W(t,\rr)\big|_{\{|\nn\rangle,|\mm\rangle\}},
  $$
where $H_\mathrm{free}\big|_{\{|\nn\rangle,|\mm\rangle\}}=E_\nn|\nn\rangle\langle\nn|+E_\mm|\mm\rangle\langle\mm|$ is the two-level atom Hamiltonian, and $H_\mathrm{em}=\left(a^\dag a+\frac12\right)\omega$, is the quantized electromagnetic mode Hamiltonian, $a$ and $a^\dag$ being its annihilation and creation operators, with $[a,a^\dag]=1$. In the electric dipole approximation, the electric field variation along the typical particle size is negligible, and the particle is assumed to be at an electric field maximum value. The dipole moment is thus oriented along the direction of the electric field, therefore $A_\nn^\mm=2|\langle\mm|\ddd|\nn\rangle|^2\frac Q{V_\mathrm{mode}}$. From the definition $A_\nn^\mm=F_\nn^\mm\!~^0\!A_\nn^\mm$ and from the spontaneous emission rate in absence of electromagnetic cavity $^0\!A_\nn^\mm=\frac{(\omega_\nn^\mm)^3}{3\pi}|\langle\mm|\ddd|\nn\rangle|^2$ \cite{Ballentine}, the Purcell factor is found to be $F_\nn^\mm=\frac{3Q}{4\pi^2}\frac{(\lambda_\nn^\mm)^3}{V_\mathrm{mode}}$.

In practice, the effective cavity mode volume $V_\mathrm{mode}$ is experimentally measured. In \cite{Vahala_2003}, a review of electromagnetic microcavities of different geometries, built by different methods, but characterized by a quality factor $Q$ and by an effective volume of the form $V_\mathrm{mode}\approx\beta\lambda^3$ is given. Here, $\lambda$ is the {\it quasi}-mode wavelength at resonance, close to the wavelength $\lambda_\nn^\mm$ associated to the Bohr angular frequency $\omega_\nn^\mm$ and $\beta$ a pure number of order unity. Then, for such cavities, the Purcell factor becomes independent from the radiative transition, and $F=\frac{3Q}{20\pi^2}\approx1.51~10^{-2}Q$ if $\beta=5$ and $Q\geq2000$. To observe the Purcell effect, {\it i.e.} to have $F\geq1$, the quality factor should be larger than a lower bound of $Q_\mathrm{min}\approx66$. Thus, in common electromagnetic cavities, the Purcell effect is generally observable and measurable.

The validity criterion for the weak coupling regime is obtained by comparing the characteristic time scale of the coupling of the two-level system to the electric field confined mode and the coupling of the electric field mode to the {\it continuum} of external electromagnetic modes. The second coupling dominates if the emitted photon during the transition escapes the two-level quantum system and the electromagnetic cavity, without being re-absorbed. The associated photon relaxation time for the radiative transition is defined as $\tau_\nn^\mm=\frac Q{\omega_\nn^\mm}$. Moreover, the Purcell effect must face the adverse working of Rabi oscillations.\footnote{For more details, one can refer to \cite{Tannoudji}.} A sufficient condition for the validity of the weak coupling is then simply
  $$
\tau_\nn^\mm\Omega_\nn^\mm\ll1,
  $$
where $\Omega_\nn^\mm=\sqrt{\frac\omega{2V_{\textrm{\tiny mode}}}}|\langle\mm|\ddd|\nn\rangle|$ is the  Rabi angular frequency of the related radiative transition. In the strong coupling regime, defined by $\tau_\nn^\mm\Omega_\nn^\mm\gg1$, only the interaction between the two-level quantum system and the confined mode is to be considered. As the dimensionless quantity $\tau_\nn^\mm\Omega_\nn^\mm$ scales as $\propto Q$, the previous condition imposes an upper bound on $Q$. In fact, the higher $Q$ is, the smaller is the resonance disagreement $|\omega-\omega_\nn^\mm|$, which is responsible for the Rabi oscillations evanescence. Therefore, Rabi oscillations can be maintained in the electromagnetic cavity, inhibiting the Purcell effect. $|\omega-\omega_\nn^\mm|$ should be then sufficiently small to insure the validity of the resonant approach, but it should not be too small not to promote Rabi oscillations, unfavorable for the Purcell effect.
\subsection{A first approach in spherical semiconductor QDs}
In a dielectric medium, the dielectric permittivity $\varepsilon$ is related to the refraction index $\eta$ by $\varepsilon=\eta^2>1$. As the Purcell effect concerns radiative transitions between any two QD eigenstates, the confined hole is assumed to sit in its ground state. Then, this approach should be reasonable in the strong confinement regime. The charge carriers are considered as uncorrelated particles, the so-called particle-in-a-sphere model. As observed in section {\bf \ref{sec_2}}, the confinement is implemented by a infinite potential well. In the following, electron tunneling is discarded, and only electronic radiative transitions involving energy levels lower than the maximum amplitude of the real finite potential step are to be considered.

Einstein spontaneous emission coefficients with or without electromagnetic cavity should be computed between two electronic eigenstates $|\psi_{lnm}\rangle$ and $|\psi_{l'n'm'}\rangle$ such that $E^\e_{ln}<E^\e_{l'n'}$ as
  $$
^0\!A_{lnm}^{l'n'm'}=\frac{64\alpha}3\eta\left(\frac{2\pi}{\lambda_{ln}^{l'n'}}\right)^3(I^{ll'}_{nn'})^2J_{ll'}^{mm'}R^2
  $$
and $A_{lnm}^{l'n'm'}=F\!~^0\!A_{lnm}^{l'n'm'}$, where $I^{ll'}_{nn'}$  and $J_{ll'}^{mm'}$ are respectively the radial and the angular integrals occurring in the matrix element $\langle\psi_{lnm}|\rr|\psi_{l'n'm'}\rangle$ complex modulus.\footnote{They can be analytically evaluated as
  \begin{eqnarray*}
J_{ll'}^{mm'}\lnd=\rnd\frac{\delta_{l'l-1}}2\left\{\frac{(l+m)(l+m-1)}{(2l+1)(2l-1)}\delta^{m'm+1}+2\frac{(l+m)(l-m)}{(2l+1)(2l-1)}\delta^{m'm}+\frac{(l-m)(l-m-1)}{(2l+1)(2l-1)}\delta^{m'm-1}\right\}
\\
\lnd\rnd~~~~+\frac{\delta_{l'l+1}}2\left\{\frac{(l+m+2)(l+m+1)}{(2l+3)(2l+1)}\delta^{m'm+1}+2\frac{(l+m+1)(l-m+1)}{(2l+3)(2l+1)}\delta^{m'm}+\frac{(l-m+2)(l-m+1)}{(2l+3)(2l+1)}\delta^{m'm-1}\right\},
  \end{eqnarray*}
and
  $$
I_{nn'}^{ll\pm1}=-\frac{k_{ln}k_{l\pm1n'}}{\left(k^2_{l\pm1n'}-k^2_{ln}\right)^2}.
  $$
The vacuum transition rates $^0\!A^{l'n'm'}_{lnm}$ are non-vanishing only if the selection rules $l'=l\pm1$ and $m'=m,m\pm1$ are satisfied. But, there is no particular selection rule for $n$ and $n'$.} Since $\lambda_{ln}^{l'n'}\propto (E^\e_{l'n'}-E^\e_{ln})^{-1}\propto R^2$, Einstein spontaneous emission coefficients are large for small QD radius. This shows a typical quantum behavior for small QDs through the spontaneous emission enhancement, even in the absence of an electromagnetic cavity.
\subsection{An example of QDs LASER device}
\begin{figure}
\caption{Three-level {\it red} QD-LASER: the pumping is realized between  a {\it ground state} level $|\mathbf g\rangle$ and a {\it excited state} $|\mathbf e\rangle$, higher than the highest level $|\mathbf i\rangle$ of the LASER transition, the {\it intermediate state}.} \label{figure_5}
    \begin{center}
\input{figure_5.pstex_t}
    \end{center}
  \end{figure}
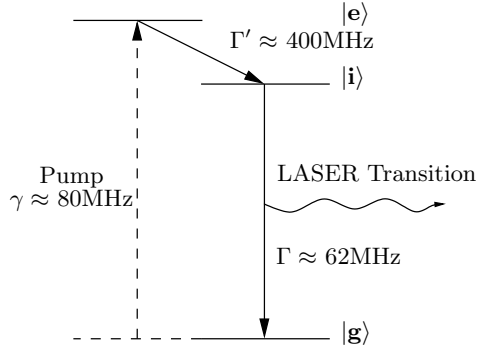

As explained before, semiconductor QDs Purcell effect could be used,  instead of real atoms, as efficient radiation emitters in  LASER devices over quite wide wavelength ranges. The general theory of LASER and the so-called population inversion is well known, see {\it e.g.} \cite{Pocholle, Grynberg}. Here, we expose the possibility of exploiting the Purcell effect to produce {\it red-light} LASER emission from spherical InAs QDs. To make contact with experimental results, we assume that $R=25$nm \cite{Lee_2002}, for which the strong confinement regime is valid. A three-level LASER is built with the previous QD states, where the transition is found at $755$nm. To have a concrete idea of the working of the LASER mechanism described by figure \ref{figure_5}, we have collected in table \ref{table_3} the relevant numerical data.

The spontaneous emission is the only phenomenon to be considered. Stimulated emission and absorption should be discarded, and non-radiative effects should be omitted. The non-radiative effects in QDs lead to energy dispersion by phonon creation, when inelastic collisions between electrons and the potential well occur. These only shorten the lifetime of excited states. So, non-radiative effects do not matter in the qualitative arguments for non-LASER transitions. In this context, we shall keep radiative effects, especially spontaneous emission effects which can be enhanced by the Purcell effect in LASER transition, and thereby initiate LASER oscillations.

  \begin{table}
\caption{Wavelengths and spontaneous emission rates of the three-level QD {\it red} LASER presented in figure \ref{figure_5} in comparison with He-Ne LASER.}
\label{table_3}
    \begin{center}
      \begin{tabular}{ccccc}
\hline\hline
LASER & Transition & Wavelength (nm) & $^0\!A_\nn^\mm$ (MHz)
\\
\hline
\multirow{2}{1.7cm}{QD-LASER} & $|\mathbf e\rangle\!\rightarrow|\mathbf i\rangle$ & 6.02~10$^3$ & 401
\\
 & $|\mathbf i\rangle\!\rightarrow|\mathbf g\rangle$ & 755 & 0.617
\\
\hline
 He-Ne & LASER & 632 & $\approx50$
\\
\hline\hline
      \end{tabular}
    \end{center}
  \end{table}

We assume that the decay $|\mathbf e\rangle\rightarrow|\mathbf i\rangle$,  governed by the relaxation rate $\Gamma'$, is faster than the decay $|\mathbf i\rangle\rightarrow|\mathbf g\rangle$ governed by the relaxation rate $\Gamma$, {\it i.e.} the intermediate state should be metastable as compared to the excited, and $\Gamma'\gg\Gamma$. In a stationary regime and in the case of weak pumping $\omega\ll\Gamma'$, the population inversion holds only for $\omega\geq\Gamma$. This implies that $\Gamma'\gg\Gamma$, which means that the excited state is almost empty, and the intermediate state is the most populated state.

As $\Gamma'\geq\!\!~^0\!A_\mathbf i^\mathbf e\approx401$MHz and $\Gamma=A^\mathbf i_\mathbf g\approx9.38Q$kHz, the  assumptions for having a red-light emitting three-level LASER are met, if the Purcell factor is about $F\approx100$, {\it i.e.} for a quality factor about $Q\approx6500$, and if we choose a pumping frequency of the order of magnitude of $\omega\approx80$MHz. In particular, the condition for obtaining the Purcell effect is met, since $\tau_\mathbf g^\mathbf i\Omega_\mathbf g^\mathbf i\approx6,3.10^{-3}\ll1$. Table \ref{table_3} suggests that spontaneous emission rates are of the same order of magnitude than those of the He-Ne LASER transition, even if He-Ne LASER are built upon a four-level system.\footnote{The mechanism of a four-level LASER is described in \cite{Pocholle}. Its principal advantage consists of, contrary to a three-level LASER, that there is no population inversion condition on the pumping frequency, only the LASER cavity losses fixes the threshold for coherent light emission.}

So, the Purcell effect coupled to the artificially tailored spectrum of the QDs allows the possibility to observe LASER emission in a QD-LASER, working with poor quality factor electromagnetic cavities. But, the present treatment, even in the strong confinement regime, based on the particle-in-a-sphere model, is intrinsically limited. To fully describe the Purcell effect, it is essential to exactly diagonalize the total Hamiltonian $H$. However, at the moment, the inclusion of electron-hole Coulomb interaction, even in the presence of an infinite confinement potential well, turns the problem into a challenge to be met.
\section{Conclusion}
In this paper, a new approach to some interesting atom-like properties of spherical semiconductor QDs is presented. It is based on an improved EMA model to which is added an effective pseudo-potential. This allows extensive analytic calculations of physical quantities yielding a better agreement with empirical data for QSE and QCSE. The Lamb shift in spherical semiconductor QDs is also calculated in this theoretical framework. It turns out to be negative and in principle observable, at least in the strong confinement regime. A {\it Gedankenexperiment}, based on a modification of the electromagnetic field vacuum fluctuations environnement for the QD, created by a Casimir configuration, is proposed. A modification of the QD Lamb shift should be observable, as compared to the one existing in free space. Finally our study also illustrates the utility of the Purcell effect, predicted for atoms, for QD-LASER emission in the visible part of the spectrum.

These wide ranging theoretical results are encouraging for further investigation of QDs structure, based on this improved EMA model. In view of a full description of phenomena involving radiative transitions between two energy levels of the QD, like Purcell effect or LASER emission, it seems relevant to develop the general theory of the confined interactive electron-hole pair states. 

Another fundamental outlook consists of investigating the description of the charged carriers confinement potential by an finite potential step instead of the infinite potential well used here, in order to fully describe QDs of particularly small radius.
\appendix
\section{Constants} \label{appendix_A}
In the following tables, we sum up all appearing constants and give their approximate values. The function $\Si(x)=\int_0^x\frac{\dd t}t\sin(t)$ denotes the standard sine integral. Table \ref{table_4} presents analytical expressions and approximate values of constants occurring in section {\bf \ref{sec_2}}.

Table \ref{table_5} presents analytical expressions and approximate values  of constants occurring in section {\bf \ref{subsec_3_2}}, when only the Coulomb potential is taken into account. We are able to provide exact expressions for all the constants occurring  in section {\bf \ref{subsec_3_3}}, when the polarization energy is also taken into account, except for $\delta'''$, $\gamma'''$ and $\gamma''''$. For these quantities, we obtain integral representations, which cannot be analytically computed at the moment. Their approximate values are evaluated numerically. As exact expressions for other constants are quite cumbersome, we give only their approximate values. Table \ref{table_6} presents approximate values for constants which appear in the polarization energy diagonal matrix element $\langle\Phi|P(\rr_\e,\rr_\h)|\Phi\rangle$, while table \ref{table_7} defines constants which appear in the polarization mean value Eq. (\ref{Phi|P(rr_e,rr_h)|Phi}) and gives their approximate values in CdS$_{0.12}$Se$_{0.88}$ microcrystals.
  \begin{table*}
\caption{Definitions, analytic expressions and approximate values of constants appearing in section {\bf \ref{sec_2}}.} \label{table_4}
    \begin{center}
      \begin{tabular}{cccccc}
\hline\hline
Name & Expression & Value & Name & Expression & Value
\\
\hline
$S$ & $\ds\Si(2\pi)-\frac{\Si(4\pi)}2$ & 0.6720 & $A$ & $\ds2-\frac S\pi$ & 1.7861
\vs{.1cm}
\\
$B_1$ & $\ds\frac23-\frac5{8\pi^2}$ & 0.6033 & $B_2$ & $\ds\frac29+\frac{13}{24\pi^2}+\frac S{2\pi^3}$ & 0.2879
\vs{.1cm}
\\
$B$ & $\ds B_1+\frac{B_2}3$ & 0.6993 & $B'$ & $AB-1$ & 0.2489
\vs{.1cm}
\\
$\delta$ & $\ds\frac{3\pi}{40}\!\left\{1-\frac2{\pi S}\right\}$ & 0.2081 &
\vs{.1cm}
\\
\hline\hline
      \end{tabular}
    \end{center}
  \end{table*}

  \begin{table*}
\caption{Definitions, analytical expressions and approximate values of constants appearing in section {\bf \ref{subsec_3_2}}.} \label{table_5}
    \begin{center}
      \begin{tabular}{cccccc}
\hline\hline
Name & Expression & Value & Name & Expression & Value
\\
\hline
$C$ & $\ds\frac13-\frac1{2\pi^2}$ & 0.2827 & $C'$ & $\ds A(B^2-C)-\frac B2$ & 0.0189
\vs{.1cm}
\\
$C'_1$ & $\ds\frac{B_1-2AC}{12}$ & -0.0339 & $C'_2$ & $\ds\frac{B_2}{18}$ & 0.0160
\vs{.1cm}
\\
$D_1$ & $\ds\frac25-\frac{13}{8\pi^2}+\frac{147}{64\pi^4}$ & 0.2589 & $D_2$ & $\ds\frac2{15}-\frac1{8\pi^2}-\frac{21}{64\pi^4}$ & 0.1173
\vs{.1cm}
\\
$D_3$ & $\ds\frac2{25}+\frac{37}{120\pi^2}-\frac{1153}{320\pi^4}-\frac{3S}{2\pi^5}$ & 0.0710 & $D$ & $\ds\frac{5D_1+10D_2+D_3}{30}$ & 0.2539
\vs{.1cm}
\\
$D'$ & $\ds\frac{3D_1+4D_2+D_3}6$ & 0.2195 & $D''$ & $\ds\frac{5D_2-D_3}{45}$ & 0.0115
\vs{.1cm}
\\
$C''$ & $\ds\frac{D'+3D''-BC}3$ & 0.0187
\vs{.1cm}
\\
\hline\hline
      \end{tabular}
    \end{center}
  \end{table*}

  \begin{table*}
\caption{Approximate values of constants appearing in $\langle\Phi|P(\rr_\textrm{\tiny e},\rr_\textrm{\tiny h})|\Phi\rangle$ in  section {\bf \ref{subsec_3_3}}.} \label{table_6}
    \begin{center}
      \begin{tabular}{cccccccccc}
\hline\hline
Name & Value & Name & Value & Name & Value & Name & Value & Name & Value
\\
\hline
$\beta'$ & 3.1144 & $\gamma'$ & 2.3218 & $\gamma'''$ & 1.3263 & $\delta'$ & 0.9973 & $\delta'''$ & 0.0533
\\
$\beta''$ & 0.7524 & $\gamma''$ & 0.5992 & $\gamma''''$ & -0.0704 & $\delta''$ & 0.2708
\\
\hline\hline
      \end{tabular}
    \end{center}
  \end{table*}

  \begin{table*}
\caption{Definitions and approximate values of constants appearing in section {\bf \ref{subsec_3_3}} for CdS$_{0.12}$Se$_{0.88}$ microcrystals.} \label{table_7}
    \begin{center}
      \begin{tabular}{cccccc}
\hline\hline
Name & Expression & Value & Name & Expression & Value
\\
\hline
$\beta(\varepsilon_\rrr)$ & $\ds\frac12\frac{\varepsilon_\rrr-1}{\varepsilon_\rrr+1}\left(~\!\beta'-2+\frac{\varepsilon_\rrr}{\varepsilon_\rrr+1}\beta''\right)$ & 0.5149 & $A(\varepsilon_\rrr)$ & $-\beta(\varepsilon_\rrr)$ & -0.5149
\vs{.1cm}
\\
$\gamma(\varepsilon_\rrr)$ & $\ds\frac12\frac{\varepsilon_\rrr-1}{\varepsilon_\rrr+1}\left(~\!\gamma'-\gamma'''+\frac{\varepsilon_\rrr}{\varepsilon_\rrr+1}(\gamma''-\gamma'''')\right)$ & 0.4594 & $B'(\varepsilon_\rrr)$ & $-\beta(\varepsilon_\rrr)B+\gamma(\varepsilon_\rrr)$ & -0.0993
\vs{.1cm}
\\
$\delta(\varepsilon_\rrr)$ & $\ds\frac{\varepsilon_\rrr-1}{\varepsilon_\rrr+1}\left(~\!\delta'+\delta'''-2C+\frac{\varepsilon_\rrr}{\varepsilon_\rrr+1}(\delta''+\delta''')\right)$ & 0.4467 & $C'(\varepsilon_\rrr)$ & $\ds-\beta(\varepsilon_\rrr)(B^2-C)+\gamma(\varepsilon_\rrr)B-\frac{\delta(\varepsilon_\rrr)}2$ & -0.0083
\vs{.1cm}
\\
$\delta_1(\varepsilon_\rrr)$ & $\ds\frac{\varepsilon_\rrr-1}{\varepsilon_\rrr+1}\left(\delta'-2C+\frac{\varepsilon_\rrr}{\varepsilon_\rrr+1}\delta''\right)$ & 0.3891 & $C'_1(\varepsilon_\rrr)$ & $\ds-\frac{\delta_1(\varepsilon_\rrr)-2\beta(\varepsilon_\rrr)C}{12}$ & -0.0082
\vs{.1cm}
\\
$\delta_2(\varepsilon_\rrr)$ & $\ds\frac{\varepsilon_\rrr-1}2\frac{2\varepsilon_\rrr+1}{(\varepsilon_\rrr+1)^2}$ & 0.5400 & $C'_2(\varepsilon_\rrr)$ & $\ds-\frac{\delta_2(\varepsilon_\rrr)}{18}$ & -0.0300
\vs{.1cm}
\\
\hline\hline
      \end{tabular}
    \end{center}
  \end{table*}

\end{document}

%% file: figure_1.pstex_t
\begin{picture}(0,0)%
\includegraphics{figure_1.pstex}%
\end{picture}%
\setlength{\unitlength}{4144sp}%
\begingroup\makeatletter\ifx\SetFigFont\undefined%
\gdef\SetFigFont#1#2#3#4#5{%
  \reset@font\fontsize{#1}{#2pt}%
  \fontfamily{#3}\fontseries{#4}\fontshape{#5}%
  \selectfont}%
\fi\endgroup%
\begin{picture}(2273,2830)(-284,-1966)
\put(  1,-1707){\makebox(0,0)[b]{\smash{{\SetFigFont{9}{10.8}{\rmdefault}{\mddefault}{\updefault}{\color[rgb]{0,0,0}0}%
}}}}
\put(-44,519){\makebox(0,0)[rb]{\smash{{\SetFigFont{9}{10.8}{\rmdefault}{\mddefault}{\updefault}{\color[rgb]{0,0,0}30}%
}}}}
\put(-44,-72){\makebox(0,0)[rb]{\smash{{\SetFigFont{9}{10.8}{\rmdefault}{\mddefault}{\updefault}{\color[rgb]{0,0,0}20}%
}}}}
\put(-44,-699){\makebox(0,0)[rb]{\smash{{\SetFigFont{9}{10.8}{\rmdefault}{\mddefault}{\updefault}{\color[rgb]{0,0,0}10}%
}}}}
\put(-44,-1290){\makebox(0,0)[rb]{\smash{{\SetFigFont{9}{10.8}{\rmdefault}{\mddefault}{\updefault}{\color[rgb]{0,0,0}0}%
}}}}
\put(-44,208){\makebox(0,0)[rb]{\smash{{\SetFigFont{9}{10.8}{\rmdefault}{\mddefault}{\updefault}{\color[rgb]{0,0,0}25}%
}}}}
\put(789,-1707){\makebox(0,0)[b]{\smash{{\SetFigFont{9}{10.8}{\rmdefault}{\mddefault}{\updefault}{\color[rgb]{0,0,0}2}%
}}}}
\put(395,-1707){\makebox(0,0)[b]{\smash{{\SetFigFont{9}{10.8}{\rmdefault}{\mddefault}{\updefault}{\color[rgb]{0,0,0}1}%
}}}}
\put(1171,-1707){\makebox(0,0)[b]{\smash{{\SetFigFont{9}{10.8}{\rmdefault}{\mddefault}{\updefault}{\color[rgb]{0,0,0}3}%
}}}}
\put(1554,-1707){\makebox(0,0)[b]{\smash{{\SetFigFont{9}{10.8}{\rmdefault}{\mddefault}{\updefault}{\color[rgb]{0,0,0}4}%
}}}}
\put(1936,-1707){\makebox(0,0)[b]{\smash{{\SetFigFont{9}{10.8}{\rmdefault}{\mddefault}{\updefault}{\color[rgb]{0,0,0}5}%
}}}}
\put(-44,-978){\makebox(0,0)[rb]{\smash{{\SetFigFont{9}{10.8}{\rmdefault}{\mddefault}{\updefault}{\color[rgb]{0,0,0}5}%
}}}}
\put(991,-1906){\makebox(0,0)[b]{\smash{{\SetFigFont{9}{10.8}{\rmdefault}{\mddefault}{\updefault}{\color[rgb]{0,0,0}$\ds\frac R{a^*}$}%
}}}}
\put(-269,-385){\makebox(0,0)[rb]{\smash{{\SetFigFont{9}{10.8}{\rmdefault}{\mddefault}{\updefault}{\color[rgb]{0,0,0}$\ds\frac{E_{\e\h}-E_\g}{E^*}$}%
}}}}
\put(-44,-385){\makebox(0,0)[rb]{\smash{{\SetFigFont{9}{10.8}{\rmdefault}{\mddefault}{\updefault}{\color[rgb]{0,0,0}15}%
}}}}
\end{picture}%

%% file: figure_2.pstex_t
\begin{picture}(0,0)%
\includegraphics{figure_2.pstex}%
\end{picture}%
\setlength{\unitlength}{4144sp}%
\begingroup\makeatletter\ifx\SetFigFont\undefined%
\gdef\SetFigFont#1#2#3#4#5{%
  \reset@font\fontsize{#1}{#2pt}%
  \fontfamily{#3}\fontseries{#4}\fontshape{#5}%
  \selectfont}%
\fi\endgroup%
\begin{picture}(4030,2757)(-329,-1796)
\put(1775,-1736){\makebox(0,0)[b]{\smash{{\SetFigFont{9}{10.8}{\rmdefault}{\mddefault}{\updefault}{\color[rgb]{0,0,0}$R$ (\AA)}%
}}}}
\put(  1,862){\makebox(0,0)[lb]{\smash{{\SetFigFont{7}{8.4}{\rmdefault}{\mddefault}{\updefault}{\color[rgb]{0,0,0}$\times10^{-2}$}%
}}}}
\put(-314,-466){\makebox(0,0)[rb]{\smash{{\SetFigFont{9}{10.8}{\rmdefault}{\mddefault}{\updefault}{\color[rgb]{0,0,0}$\Delta E^\mathrm{strong}_\mathrm{Stark}$ (meV)}%
}}}}
\put( -1,-1574){\makebox(0,0)[b]{\smash{{\SetFigFont{9}{10.8}{\rmdefault}{\mddefault}{\updefault} 0}}}}
\put(564,-1574){\makebox(0,0)[b]{\smash{{\SetFigFont{9}{10.8}{\rmdefault}{\mddefault}{\updefault} 10}}}}
\put(1170,-1574){\makebox(0,0)[b]{\smash{{\SetFigFont{9}{10.8}{\rmdefault}{\mddefault}{\updefault} 20}}}}
\put(1775,-1574){\makebox(0,0)[b]{\smash{{\SetFigFont{9}{10.8}{\rmdefault}{\mddefault}{\updefault} 30}}}}
\put(2382,-1574){\makebox(0,0)[b]{\smash{{\SetFigFont{9}{10.8}{\rmdefault}{\mddefault}{\updefault} 40}}}}
\put(2987,-1574){\makebox(0,0)[b]{\smash{{\SetFigFont{9}{10.8}{\rmdefault}{\mddefault}{\updefault} 50}}}}
\put(3594,-1574){\makebox(0,0)[b]{\smash{{\SetFigFont{9}{10.8}{\rmdefault}{\mddefault}{\updefault} 60}}}}
\put(-44,-976){\makebox(0,0)[rb]{\smash{{\SetFigFont{9}{10.8}{\rmdefault}{\mddefault}{\updefault}-1.5}}}}
\put(-44,-465){\makebox(0,0)[rb]{\smash{{\SetFigFont{9}{10.8}{\rmdefault}{\mddefault}{\updefault}-1}}}}
\put(-44, -2){\makebox(0,0)[rb]{\smash{{\SetFigFont{9}{10.8}{\rmdefault}{\mddefault}{\updefault}-0.5}}}}
\put(-44,509){\makebox(0,0)[rb]{\smash{{\SetFigFont{9}{10.8}{\rmdefault}{\mddefault}{\updefault}0}}}}
\put(-44,-1484){\makebox(0,0)[rb]{\smash{{\SetFigFont{9}{10.8}{\rmdefault}{\mddefault}{\updefault}-2}}}}
\end{picture}%

%% file: figure_3a.pstex_t
\begin{picture}(0,0)%
\includegraphics{figure_3a.pstex}%
\end{picture}%
\setlength{\unitlength}{4144sp}%
\begingroup\makeatletter\ifx\SetFigFont\undefined%
\gdef\SetFigFont#1#2#3#4#5{%
  \reset@font\fontsize{#1}{#2pt}%
  \fontfamily{#3}\fontseries{#4}\fontshape{#5}%
  \selectfont}%
\fi\endgroup%
\begin{picture}(4215,2758)(-284,-1796)
\put(1967,-1736){\makebox(0,0)[b]{\smash{{\SetFigFont{9}{10.8}{\rmdefault}{\mddefault}{\updefault}{\color[rgb]{0,0,0}$R$ (\AA)}%
}}}}
\put(  3,839){\makebox(0,0)[lb]{\smash{{\SetFigFont{7}{8.4}{\rmdefault}{\mddefault}{\updefault}{\color[rgb]{0,0,0}$\times10^{-5}$}%
}}}}
\put(-269,-412){\makebox(0,0)[rb]{\smash{{\SetFigFont{9}{10.8}{\rmdefault}{\mddefault}{\updefault}{\color[rgb]{0,0,0}$\ds\frac{\Delta E^\mathrm{strong}_\mathrm{Lamb}}{E_{\e\h}}$}%
}}}}
\put(3916,839){\makebox(0,0)[lb]{\smash{{\SetFigFont{9}{10.8}{\rmdefault}{\mddefault}{\updefault}{\color[rgb]{0,0,0}{\bf a.}}%
}}}}
\put(-17,-1082){\makebox(0,0)[rb]{\smash{{\SetFigFont{9}{10.8}{\rmdefault}{\mddefault}{\updefault}-5}}}}
\put(-17,-746){\makebox(0,0)[rb]{\smash{{\SetFigFont{9}{10.8}{\rmdefault}{\mddefault}{\updefault}-4}}}}
\put(-17,-412){\makebox(0,0)[rb]{\smash{{\SetFigFont{9}{10.8}{\rmdefault}{\mddefault}{\updefault}-3}}}}
\put(-17,-77){\makebox(0,0)[rb]{\smash{{\SetFigFont{9}{10.8}{\rmdefault}{\mddefault}{\updefault}-2}}}}
\put(-17,257){\makebox(0,0)[rb]{\smash{{\SetFigFont{9}{10.8}{\rmdefault}{\mddefault}{\updefault}-1}}}}
\put(-17,590){\makebox(0,0)[rb]{\smash{{\SetFigFont{9}{10.8}{\rmdefault}{\mddefault}{\updefault} 0}}}}
\put(247,-1675){\makebox(0,0)[b]{\smash{{\SetFigFont{9}{10.8}{\rmdefault}{\mddefault}{\updefault} 0.1}}}}
\put(1601,-1675){\makebox(0,0)[b]{\smash{{\SetFigFont{9}{10.8}{\rmdefault}{\mddefault}{\updefault} 1}}}}
\put(2960,-1675){\makebox(0,0)[b]{\smash{{\SetFigFont{9}{10.8}{\rmdefault}{\mddefault}{\updefault} 10}}}}
\put(-17,-1415){\makebox(0,0)[rb]{\smash{{\SetFigFont{9}{10.8}{\rmdefault}{\mddefault}{\updefault}-6}}}}
\end{picture}%

%% file: figure_3b.pstex_t
\begin{picture}(0,0)%
\includegraphics{figure_3b.pstex}%
\end{picture}%
\setlength{\unitlength}{4144sp}%
\begingroup\makeatletter\ifx\SetFigFont\undefined%
\gdef\SetFigFont#1#2#3#4#5{%
  \reset@font\fontsize{#1}{#2pt}%
  \fontfamily{#3}\fontseries{#4}\fontshape{#5}%
  \selectfont}%
\fi\endgroup%
\begin{picture}(4535,3144)(-244,-2155)
\put(-14,879){\makebox(0,0)[lb]{\smash{{\SetFigFont{6}{7.2}{\rmdefault}{\mddefault}{\updefault}{\color[rgb]{0,0,0}$\times10^{-5}$}%
}}}}
\put(2150,-2098){\makebox(0,0)[b]{\smash{{\SetFigFont{9}{10.8}{\rmdefault}{\mddefault}{\updefault}{\color[rgb]{0,0,0}$R$ (\AA)}%
}}}}
\put(-229,-493){\makebox(0,0)[rb]{\smash{{\SetFigFont{9}{10.8}{\rmdefault}{\mddefault}{\updefault}{\color[rgb]{0,0,0}$\ds\frac{\Delta E^\mathrm{strong}_\mathrm{Lamb}}{E_{\e\h}}$}%
}}}}
\put(4276,865){\makebox(0,0)[lb]{\smash{{\SetFigFont{9}{10.8}{\rmdefault}{\mddefault}{\updefault}{\color[rgb]{0,0,0}{\bf b.}}%
}}}}
\put(-35,-1502){\makebox(0,0)[rb]{\smash{{\SetFigFont{9}{10.8}{\rmdefault}{\mddefault}{\updefault}-3}}}}
\put(-35,-193){\makebox(0,0)[rb]{\smash{{\SetFigFont{9}{10.8}{\rmdefault}{\mddefault}{\updefault}-1}}}}
\put(-35,460){\makebox(0,0)[rb]{\smash{{\SetFigFont{9}{10.8}{\rmdefault}{\mddefault}{\updefault} 0}}}}
\put(668,-1933){\makebox(0,0)[b]{\smash{{\SetFigFont{9}{10.8}{\rmdefault}{\mddefault}{\updefault} 1}}}}
\put(3541,-1933){\makebox(0,0)[b]{\smash{{\SetFigFont{9}{10.8}{\rmdefault}{\mddefault}{\updefault} 100}}}}
\put(2106,-1933){\makebox(0,0)[b]{\smash{{\SetFigFont{9}{10.8}{\rmdefault}{\mddefault}{\updefault} 10}}}}
\put(-35,-849){\makebox(0,0)[rb]{\smash{{\SetFigFont{9}{10.8}{\rmdefault}{\mddefault}{\updefault}-2}}}}
\end{picture}%

%% file: figure_4.pstex_t
\begin{picture}(0,0)%
\includegraphics{figure_4.pstex}%
\end{picture}%
\setlength{\unitlength}{4144sp}%
\begingroup\makeatletter\ifx\SetFigFont\undefined%
\gdef\SetFigFont#1#2#3#4#5{%
  \reset@font\fontsize{#1}{#2pt}%
  \fontfamily{#3}\fontseries{#4}\fontshape{#5}%
  \selectfont}%
\fi\endgroup%
\begin{picture}(5400,3467)(1291,-3901)
\put(4501,-3841){\makebox(0,0)[b]{\smash{{\SetFigFont{10}{12.0}{\rmdefault}{\mddefault}{\updefault}{\color[rgb]{0,0,0}$R$ (\AA)}%
}}}}
\put(1306,-2029){\makebox(0,0)[rb]{\smash{{\SetFigFont{10}{12.0}{\rmdefault}{\mddefault}{\updefault}{\color[rgb]{0,0,0}$\ds\frac{\Delta E_\mathrm{Casimir}^\e(d)}{\Delta E_\e}$}%
}}}}
\put(1750,-2639){\makebox(0,0)[rb]{\smash{{\SetFigFont{10}{12.0}{\rmdefault}{\mddefault}{\updefault}0.05}}}}
\put(1750,-1420){\makebox(0,0)[rb]{\smash{{\SetFigFont{10}{12.0}{\rmdefault}{\mddefault}{\updefault}0.15}}}}
\put(1750,-810){\makebox(0,0)[rb]{\smash{{\SetFigFont{10}{12.0}{\rmdefault}{\mddefault}{\updefault}0.2}}}}
\put(1822,-3672){\makebox(0,0)[b]{\smash{{\SetFigFont{10}{12.0}{\rmdefault}{\mddefault}{\updefault}0}}}}
\put(2901,-3672){\makebox(0,0)[b]{\smash{{\SetFigFont{10}{12.0}{\rmdefault}{\mddefault}{\updefault}5}}}}
\put(3981,-3672){\makebox(0,0)[b]{\smash{{\SetFigFont{10}{12.0}{\rmdefault}{\mddefault}{\updefault}10}}}}
\put(5060,-3672){\makebox(0,0)[b]{\smash{{\SetFigFont{10}{12.0}{\rmdefault}{\mddefault}{\updefault}15}}}}
\put(6139,-3672){\makebox(0,0)[b]{\smash{{\SetFigFont{10}{12.0}{\rmdefault}{\mddefault}{\updefault}20}}}}
\put(1750,-2029){\makebox(0,0)[rb]{\smash{{\SetFigFont{10}{12.0}{\rmdefault}{\mddefault}{\updefault}0.1}}}}
\put(1750,-3249){\makebox(0,0)[rb]{\smash{{\SetFigFont{10}{12.0}{\rmdefault}{\mddefault}{\updefault}0}}}}
\end{picture}%

%% file: figure_5.pstex_t
\begin{picture}(0,0)%
\includegraphics{figure_5.pstex}%
\end{picture}%
\setlength{\unitlength}{4144sp}%
\begingroup\makeatletter\ifx\SetFigFont\undefined%
\gdef\SetFigFont#1#2#3#4#5{%
  \reset@font\fontsize{#1}{#2pt}%
  \fontfamily{#3}\fontseries{#4}\fontshape{#5}%
  \selectfont}%
\fi\endgroup%
\begin{picture}(2628,2101)(-63,-2582)
\put(1747,-997){\makebox(0,0)[lb]{\smash{{\SetFigFont{9}{10.8}{\rmdefault}{\mddefault}{\updefault}{\color[rgb]{0,0,0}$|\mathbf i\rangle$}%
}}}}
\put(1747,-616){\makebox(0,0)[lb]{\smash{{\SetFigFont{9}{10.8}{\rmdefault}{\mddefault}{\updefault}{\color[rgb]{0,0,0}$|\mathbf e\rangle$}%
}}}}
\put(1366,-1577){\makebox(0,0)[lb]{\smash{{\SetFigFont{9}{10.8}{\rmdefault}{\mddefault}{\updefault}{\color[rgb]{0,0,0}LASER Transition}%
}}}}
\put(1359,-2073){\makebox(0,0)[lb]{\smash{{\SetFigFont{9}{10.8}{\rmdefault}{\mddefault}{\updefault}{\color[rgb]{0,0,0}$\Gamma\approx$ 62MHz}%
}}}}
\put(1747,-2522){\makebox(0,0)[lb]{\smash{{\SetFigFont{9}{10.8}{\rmdefault}{\mddefault}{\updefault}{\color[rgb]{0,0,0}$|\mathbf g\rangle$}%
}}}}
\put(136,-1726){\makebox(0,0)[b]{\smash{{\SetFigFont{9}{10.8}{\rmdefault}{\mddefault}{\updefault}{\color[rgb]{0,0,0}$\gamma\approx$ 80MHz}%
}}}}
\put(136,-1591){\makebox(0,0)[b]{\smash{{\SetFigFont{9}{10.8}{\rmdefault}{\mddefault}{\updefault}{\color[rgb]{0,0,0}Pump}%
}}}}
\put(1081,-781){\makebox(0,0)[lb]{\smash{{\SetFigFont{9}{10.8}{\rmdefault}{\mddefault}{\updefault}{\color[rgb]{0,0,0}$\Gamma'\approx$ 400MHz}%
}}}}
\end{picture}%